\documentclass{aa}  
\usepackage{natbib}
\bibpunct{(}{)}{;}{a}{}{,} 
\usepackage{graphicx}
\usepackage{txfonts,textcomp}
\usepackage{color}

\begin{document}

   \title{Multiwavelength observation of 1A 0535+262=HD 245770 from 2010 to 2021}

   \subtitle{}

   \author{{Wei Liu}
          \inst{1}
          \and
          Jingzhi Yan \inst{1}
          \and
          Guangcheng Xiao \inst{1}
          \and
          Xiukun Li \inst{1,2}
          \and
          Bo Gao \inst{1,2}
          \and
          Qingzhong Liu \inst{1}
          }

   \institute{Key Laboratory of Dark Matter and Space Astronomy, Purple Mountain Observatory, Chinese Academy of Sciences, Nanjing, 210023, P.R. China
         \and
             School of Astronomy and Space Science, University of Science and Technology of China, Hefei, 230026, P.R. China\\
             \email{a010110288@163.com,jzyan@pmo.ac.cn}              
             }

   \date{Received August 2, 2023; accepted September 21, 2023}

 
  \abstract
   {1A 0535+262 is a high-mass X-ray binary that went into a giant X-ray outburst in 2020. During this event, the X-ray luminosity reached the highest value measured over the last 30 years.}
   {Our aim is to study the long-term variability of 1A 0535+262 before and after the 2020 major X-ray outburst and to uncover the mechanism that led to the X-ray outburst.}
   {We used the long-term photometric light curve and the equivalent widths of the H$\alpha$ and He I $\lambda$6678 lines to monitor the state of the Be star's circumstellar disk. The H$\alpha$ line profiles show evidence for V/R variability, which we revealed by fitting the H$\alpha$ spectral line profiles with two Gaussian functions. In addition, we divided our data into four periods according to the intensity of the X-ray, optical, and infrared emission.}
   {The H$\alpha$ line profiles show single-peaked profiles in most cases. This is consistent with the previously reported orbital inclination of \emph{i} = $37^{\circ} \pm 2^{\circ}$. Unlike the H$\alpha$ lines, the He I $\lambda6678$ lines show a maximal intensity in October 2020, which is one month before the giant X-ray outburst in 2020. Based on the behavior of the equivalent widths of the H$\alpha$ and He I $\lambda6678$ lines, and the {\it V}-band magnitude, we find two mass ejection processes from the Be star to the Be disk on MJD 55820 and MJD 56600. The V/R quasi-period is about two\, years during 2011--2015, which is different from 1994 to 1995. Furthermore, the periods I$\to$II$\to$III$\to$IV in the $(B-V)$ color index versus $V$-band magnitude diagram constitute a cycle. From the behavior of the V/R ratio of H$\alpha$ lines, and the variability of the $V$ band, we believe that the precession of the density perturbation inside the disk is retrograde. 
   }
   {}

   \keywords{stars: emission-line, Be -- binaries: close -- X-rays: binaries -- stars: individual: 1A 0535+262 -- stars: neutron
               }

   \maketitle


\section{Introduction} \label{sec:intro}

1A 0535+262 is a transient Be/X-ray binary found by Ariel V in 1975 \citep{1975Natur.256..628R}. Its optical identification, HD 245770/V725 Tau, is an early-type emission-line star (O9.7IIIe) \citep{1980A&AS...40..289G, 1992SSRv...59....1G}. The orbital period is $\backsim$111 days in an eccentric orbit (e = 0.47 $\pm$ 0.02) and the pulse period of the neutron star (NS) is $\backsim$103 s \citep{1994IAUC.5931....1F, 1996ApJ...459..288F}. The distance of 1A 0535+262 is $1.9^{+0.1}_{-0.1}$ kpc as provided by the \textit{Gaia} catalog \citep{2020yCat.1350....0G}. Before \textit{Gaia}, \citet{1998MNRAS.297L...5S} used a spectral class of B0III and a reddening E(B - V) = 0.75 to obtain a distance of 2 kpc to the system. The orbital inclination \emph{i}\footnote{where \emph{i} is the inclination angle of the orbit of the disk with respect to the plane of the sky.} of the binary system is $37^{\circ} \pm 2^{\circ}$ \citep{2007A&A...475..651G}.

For 1A 0535+262, the long-term variabilities of optical/infrared photometry and optical spectral can be found in previous publications from 2010 to 2020 \citep[see e.g.,][]{2012ApJ...744...37Y, 2012ApJ...754...20C, 2013PASJ...65...83M, 2017ARep...61..983T, 2020Ap.....63..367K}. 
\citet{2012ApJ...754...20C} and \citet{2012ApJ...744...37Y} found an anti-correlation between the equivalent widths of the H$\alpha$ lines (hereafter EW(H$\alpha$)) and the {\it V} magnitudes around the giant outburst. \citet{2012ApJ...744...37Y} interpreted the decrease in optical brightness during the giant outbursts as being a result of mass ejection, which formed a tenuous region in the inner part of the Be disk. Furthermore, \citet{2012ApJ...754...20C} and \citet{2012ApJ...744...37Y} found that the optical {\it V}-band light curve of 1A 0535+262 indicated that each giant X-ray outburst occurred during a phase of decreased optical brightness, while the H$\alpha$ lines showed a strong emission.  
\citet{2020Ap.....63..367K} observd the same phenomenon as \citet{2012ApJ...754...20C} and \citet{2012ApJ...744...37Y}, and predicted the occurrence of the X-ray giant outburst in 2020.
\citet{2017ARep...61..983T} found that the IR brightness and color minima of 1A 0535+262/V725 Tau in 2003 to 2011 coincided with episodes of activity of the X-ray component of the binary.

1A 0535+262 underwent a giant X-ray outburst in November and December 2020. Its X-ray flux reached a record value in the history of 1A 0535+262 of $\backsim$12.5 Crab on November 19, 2020 (MJD 59172)  \citep{2022MNRAS.511.1121M}. The giant X-ray outburst lasted for nearly seven weeks and was observed by {\it NuSTAR}, {\it Swift}, {\it NICER}, {\it Fermi}, and {\it Chandra} at multiple wavelengths \citep{2022MNRAS.511.1121M}. 
During this giant X-ray outburst, \citet{2022MNRAS.512.1141H} reported a $\gamma$-ray emission excess at the position of 1A 0535+262, which shows a weak correlation between $\gamma$-ray flux and X-ray activity, suggesting that NS accretion could be responsible for the $\gamma$-ray emission. 
\citet{2020ATel14193....1V} observed 1A 0535+262 with the Karl G. Jansky Very Large Array (VLA) on November 10 and 15, 2020, in order to search for a radio counterpart. These authors observed at C band, centered at 6 GHz, and with a bandwidth of 4 GHz. The flux density increased from 13 $\pm$ 4 $\mu$Jy to 39 $\pm$ 4 $\mu$Jy. The coupled increase in X-ray and radio flux of 1A 0535+262 shows that the radio emission is directly related to the accretion state at that time, which is similar to the behavior seen in the transient Be/X-ray binary Swift J0243.6+6124 \citep{2018Natur.562..233V}.

In this paper together with the previous published data and several astrophysical databases, we present spectroscopic and photometric results in infrared and optical bands of 1A 0535+262 from the last 10 years, including the infrared 3.4$\mu$m and 4.6$\mu$m bands, the {\it V} band, and the H$\alpha$ and He I $\lambda6678$ emission lines. In addition, the X-ray light curves and the spin-frequency history of the NS are presented in this work as a reference. We mainly discuss the optical variabilities of the binary system and uncover the relationship between the Be disk evolution and the X-ray activities.

\section{Observations} \label{sec:Obser}

\subsection{Optical spectroscopy} 

Optical spectroscopic observations were obtained from two telescopes at two different observatories: most observations from the Xinglong Station of National Astronomical Observatories in Hebei province (China) were made with the spectrometer OptoMechanics Research (OMR) or BAO Faint Object Spectrograph and Camera (BFOSC) on the 2.16 m telescope; other data from the Lijiang station of Yunnan Astronomical Observatory in Yunnan province (China) came from the spectrometer Yunnan Faint Object Spectrograph and Camera (YFOSC) on the 2.4 m telescope. 

The OMR is equipped with a 1024 $\times$ 1024 (24 micron) pixels TK1024AB2 CCD. The OMR Grism 4 is 1200 lp $\rm mm^{-1}$, giving a nominal dispersion of 1.0 \AA\ pixel$^{-1}$ \citep{2016PASP..128k5005F}. The spectral resolution is about 2.73 \AA, and the spectra cover the wavelength range of 5500--6900 \AA, and so the spectral resolving power is about 2271.
The BFOSC is equipped with a 2048 $\times$ 2048 (15 micron) pixels Loral Lick 3 CCD. The nominal dispersion of the BFOSC Grism 8 is 1.79 \AA\ pixel$^{-1}$ \citep{2016PASP..128k5005F}. The spectral resolution is about 2.38 \AA, and the spectra cover the wavelength range of 5800--8280 \AA, and so the spectral resolving power is about 2958.
The YFOSC is equipped with a 2k $\times$ 4k (13.5 micron) pixels E2V 42-90 CCD. The nominal dispersion of the YFOSC Grism 8 is 1.47 \AA\ pixel$^{-1}$. The spectral resolution is about 10.32 \AA, and the spectra cover the wavelength range of 4970--9830 \AA, and so the spectral resolving power is about 717.

We use the IRAF\footnote{IRAF is distributed by NOAO, which is operated by the Association of Universities for Research in Astronomy, Inc., in cooperation with the National Science Foundation.} software package to reduce all the spectra, including bias-subtracted correction, flat-field correction, and cosmic-ray subtraction. We take the helium-argon spectra to obtain the pixel--wavelength relationship. In order to ensure the consistency of spectral processing, all spectra have been normalized to adjacent continua. EW(H$\alpha$) is measured by selecting two points, one on either side of the emission line, and we use the program provided in IRAF to integrate the flux relative to the straight line between the two points. The measurements were repeated five times for each spectrum, and the error was estimated based on the distribution of the obtained values. The typical error of EW(H$\alpha$) is within 5\%. The value of the error is determined by the quality of the continuum. The equivalent widths of the He I $\lambda6678$ lines (hereafter EW(He I $\lambda6678$) for short) are measured by the same method as H$\alpha$ lines.

Table~\ref{table1} gives the summaries of the spectroscopic observations. This table contains instrumental information and the results of the spectral analysis: the equivalent widths of the H$\alpha$ lines and the He I $\lambda6678$ lines. 

In addition, we also adopt the optical spectroscopic data for EW(H$\alpha$) and EW(He I $\lambda6678$) from \citet{2016ATel.8633....1S, 2013PASJ...65...83M, 2012ApJ...754...20C}, and plot them in Fig.~\ref{EW}. The spectrum datum on February 20, 2020, plotted with our label is from \citet{2020Ap.....63..367K}.

The measurements of EW(H$\alpha$) by \citet{2013PASJ...65...83M} provided much lower values than ours and others. \citet{2012ApJ...754...20C} mentioned that most likely these differences are not real but the product of two different instrument systems (spectral resolution and continuum determinations). In order to correct for these differences, we multiplied EW(H$\alpha$) and EW(He I $\lambda6678$) of \citet{2013PASJ...65...83M} by 1.38.

\subsection{Optical photometry}

Optical photometric observations are obtained from three telescopes at two different observatories: the data from the Xinglong Station of National Astronomical Observatories were made with the 80 cm Tsinghua-NAOC Telescope (TNT) and the 60 cm telescope; the data from the Lijiang station of Yunnan Astronomical Observatory come from the 2.4 m telescope.

The 80 cm telescope is an equatorial-mounted Cassegrain system with a focal ratio of f/10; it was made by AstroOptik, funded by Tsinghua University in 2004 and jointly operated with NAOC, which is equipped with the PI VersArray 1300B LN 1340 × 1300 thin, back-illuminated CCD with a 20 $\mu$m pixel size \citep{2012RAA....12.1585H}.  In this configuration, the plate scale is 0.52" pixel$^{-1}$, which provides a field of view of 11.5 × 11.2 $\rm arcmin^{2}$.
The 60 cm telescope is an equatorial-mounted system with a focal ratio of f/4.23, which is equipped with the Andor DU934P-BEX2-DD 1024 × 1024 CCD, which is providing a field of view of 18 × 18 $\rm arcmin^{2}$.
The 2.4 m telescope is an altazimuth-mounted Cassegrain system with a focal ratio of f/8,  which is equipped with the E2V CCD42-90  2k × 2k thin, back-illuminated, deep-depletion CCD with a 13.5 $\mu$m pixel size. In this configuration, the plate scale is 0.28" pixel$^{-1}$, which is providing a field of view of 9.6 × 9.6 $\rm arcmin^{2}$.

In all three telescopes, 1A 0535+262 was observed through the standard Johnson-Cousins {\it B, V, R,} and {\it I} filters. The photometric data reduction was performed using standard routines and aperture photometry packages (zphot) in IRAF, including bias subtraction and flat-field correction. In order to derive the variation of the optical brightness, we selected the reference star C2 ($\alpha$: 05 39 09.5, $\delta$: +26 22 25, J2000, according to \citet{2015A&A...574A..33R}, the average magnitudes of the reference star are {\it B} = 10.211 ± 0.014, {\it V} = 10.081 ± 0.010, {\it R} = 9.994 ± 0.010, and {\it I} = 9.878 ± 0.018) in the field of view of 1A 0535+262 to derive its differential magnitudes. The photometric magnitudes are given in Table \ref{sec:table2}.

In order to study the long-term optical variability of the source, we use the public optical photometric data from the \emph{ASAS–SN}\footnote{https://asas-sn.osu.edu/variables/f3915727-695f-5fb2-b58a-61a270b89470} Variable Stars Database \citep{2014ApJ...788...48S, 2019MNRAS.485..961J}. 
We also make use of the data from the international database of the American Association of Variable Star Observers (\emph{AAVSO})\footnote{https://app.aavso.org/webobs/results/?star=000-BBJ-814\&num\_results=200}. We also make use of the data from \emph{INTEGRAL}-OMC\footnote{Based on observations with \emph{INTEGRAL}, an ESA project with instruments and science data center funded by ESA member states (especially the PI countries: Denmark, France, Germany, Italy, Switzerland, Spain) and with the participation of Russia and the USA.}\footnote{https://sdc.cab.inta-csic.es/omc/secure/form\_busqueda.jsp}.
In addition, we adopt the data from \citet{2020Ap.....63..367K, 2012ApJ...754...20C}. 
In order to correct the systematic errors between our observations and old data from archives and the literature, we add the brightness of our {\it V}-band data $\Delta$V = 0.07 mag in Fig.~\ref{EW}. The brightness derived from \emph{ASAS–SN} is systematically dimmer than that derived from other archives, and so we add the brightness lost from the total observed flux. The applied correction is $\Delta$V = 0.33 mag. 
Only the photometric observations made with the Johnson {\it V}-band filter are adopted and plotted in the fourth panel of Fig.~\ref{EW}.

\subsection{NEOWISE photometry}

We make use of the light curves in the W1 (3.4 $\mu$m) and W2 (4.6 $\mu$m) bands provided by the Wide-field Infrared Survey Explorer \citep[\emph{WISE},][]{2010AJ....140.1868W} and \emph{NEOWISE} \citep{2011ApJ...731...53M} project  through the \emph{IRSA} viewer\footnote{https://irsa.ipac.caltech.edu/irsaviewer/?\_\_action=layout.show\\DropDown\&view=MultiTableSearchCmd}, and plot them in the fifth panel of Fig.~\ref{EW}.

\subsection{X-Ray observations}

The Burst Alert Telescope (BAT)\footnote{https://swift.gsfc.nasa.gov/results/transients/1A0535p262/} on board \emph{Swift} \citep{2013ApJS..209...14K}, the Monitor of All-sky X-ray Image (\emph{MAXI}),\footnote{http://maxi.riken.jp/star\_data/J0538+263/J0538+263.html} and the Gamma-ray Burst Monitor (GBM)\footnote{https://gammaray.nsstc.nasa.gov/gbm/science/pulsars/\\lightcurves/a0535.html} on board \emph{Fermi} \citep{2009ApJ...702..791M} have monitored 1A 0535+262 in the hard X-ray energy band (15--50 keV with BAT, 2--20 keV with \emph{MAXI}, and 12--50 keV with GBM) for a long time. Two type-II X-ray outbursts and several type-I outbursts were detected between 2010 and 2021. The X-ray band light curves from BAT (15--50 keV) and \emph{MAXI} (2--20 keV) are plotted in the first panel of Fig.~\ref{EW}. The spin-frequency history measured with GBM is plotted in the second panel of Fig.~\ref{EW}.

\section{Results} \label{sec:Resul}


\subsection{H$\alpha$  and He I $\lambda6678$ line profiles during our 2010--2021 observations}

\begin{figure}
   \centering
   \includegraphics[bb=50 100 600 1200,width=0.7\hsize]{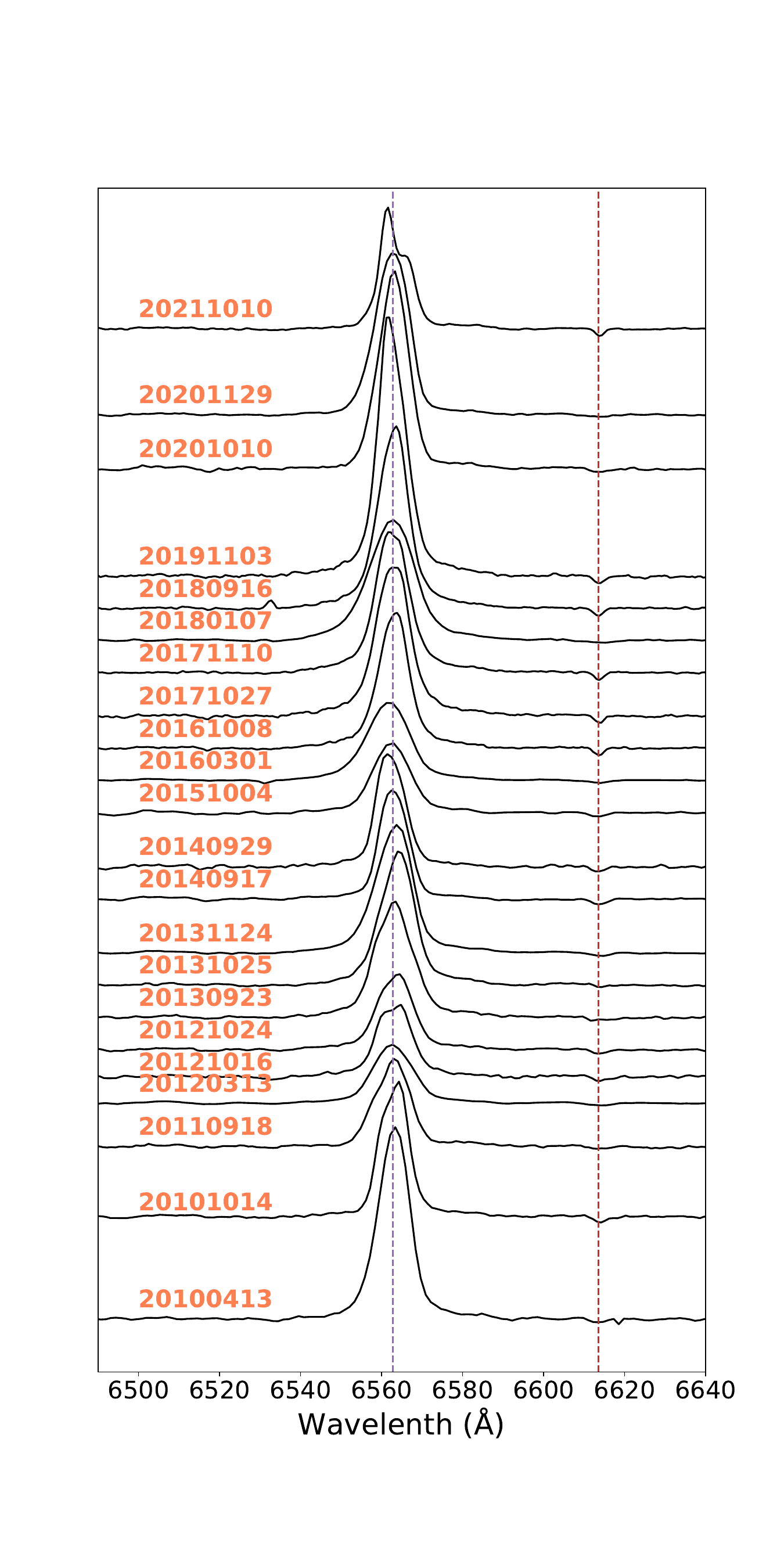}
      \caption{Evolution of H$\alpha$ line profiles between 2010 and 2021. The vertical lines mark the rest wavelength of the H$\alpha$ line and the diffuse interstellar bands at 6613 \AA. All spectra have been normalized with adjacent continua.
              }
         \label{profile_Ha}
   \end{figure}

\begin{figure}
   \centering
   \includegraphics[bb=50 50 800 850,width=\hsize]{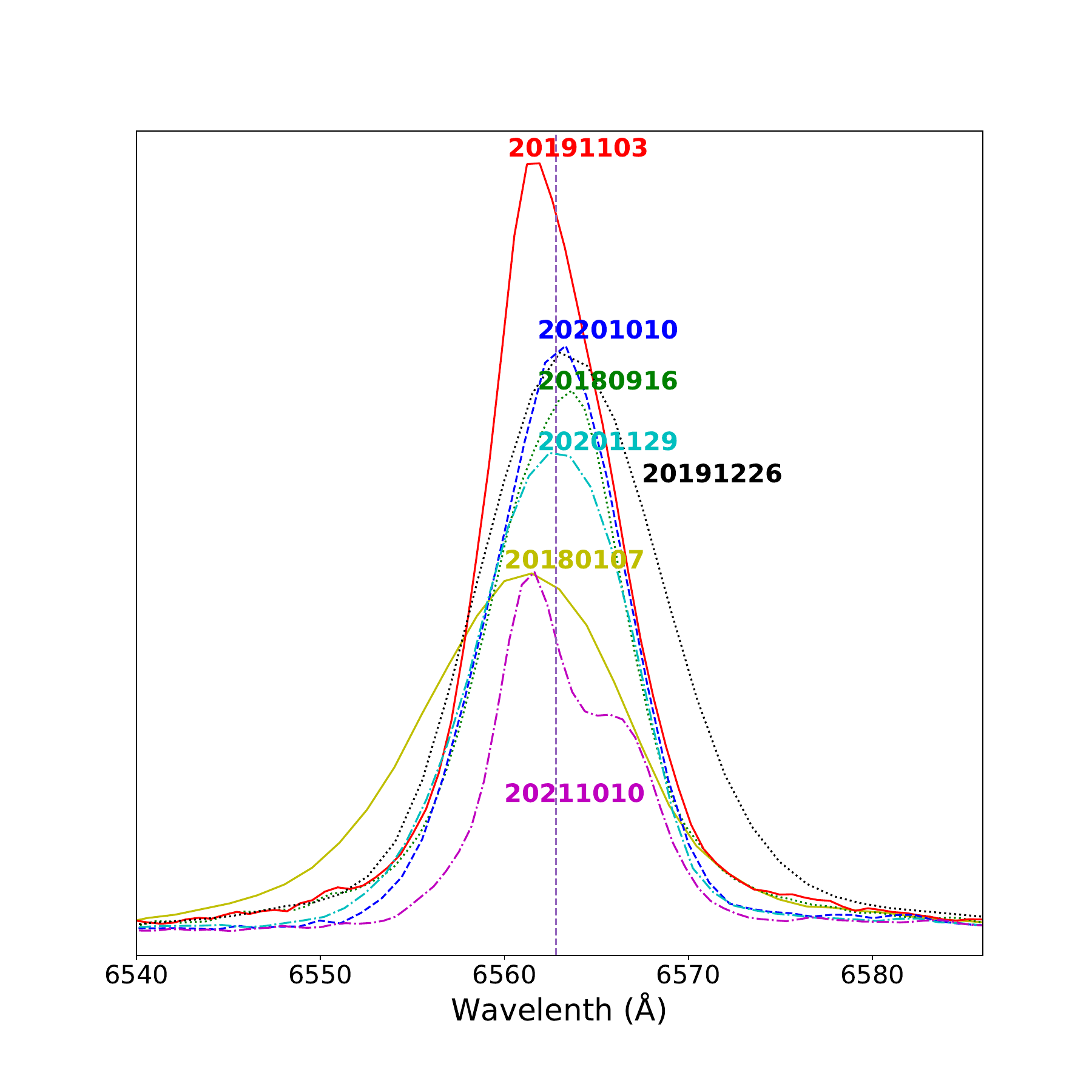}
      \caption{Evolution of H$\alpha$ line profiles between 2018 and 2021. The vertical lines mark the rest wavelength of the H$\alpha$ line. 
              }
         \label{prof_Ha2018}
   \end{figure}

\begin{figure}
   \centering
   \includegraphics[bb=50 50 800 750,width=\hsize]{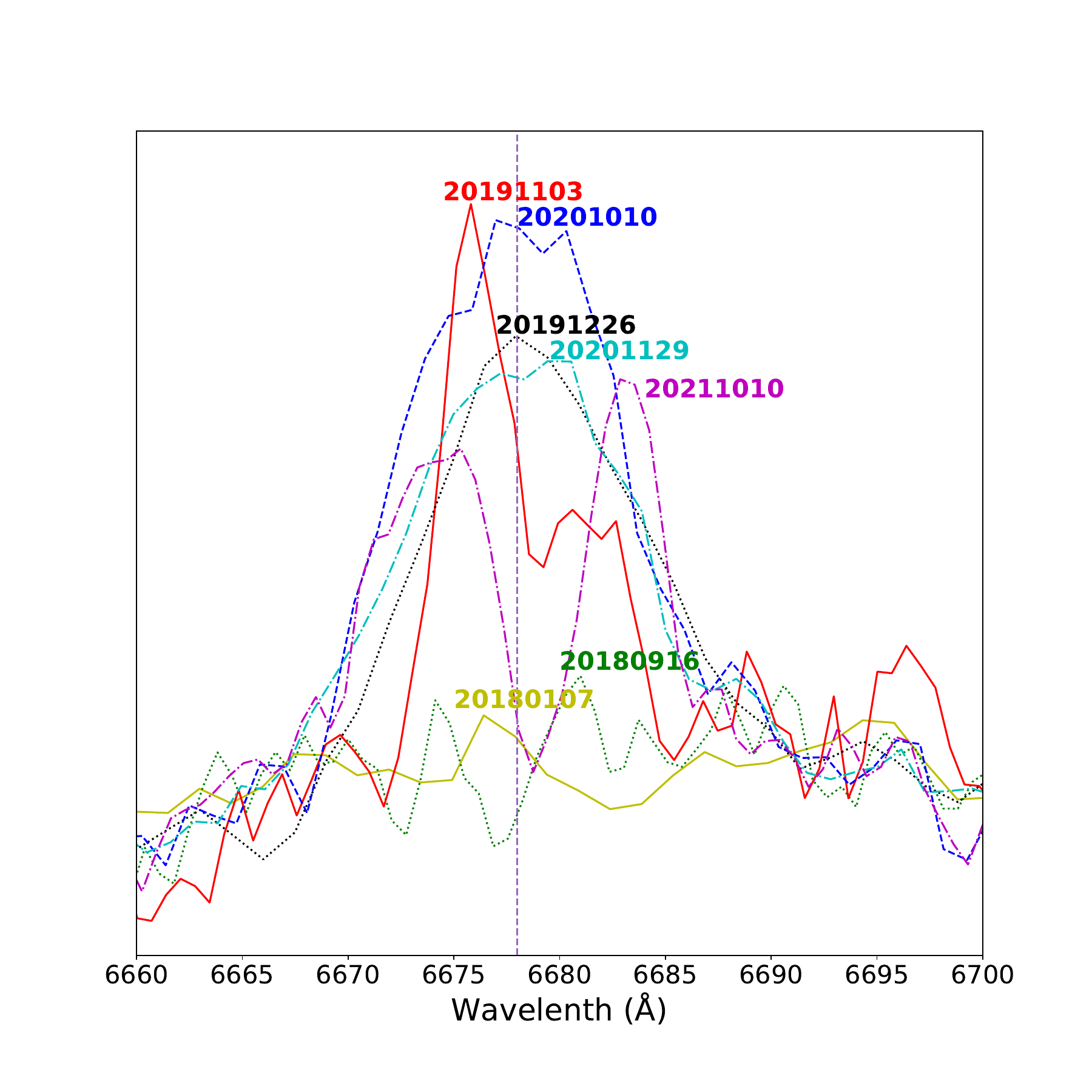}
      \caption{Evolution of He I $\lambda6678$ line profiles between 2018 and 2021. The vertical lines mark the rest wavelength of the He I $\lambda6678$ line. 
              }
         \label{prof_HeI}
   \end{figure}

In order to study the variations of the H$\alpha$ and He I $\lambda6678$ line profiles before, during, and after the 2020 giant X-ray outburst, we plot the typical spectra covering H$\alpha$ and He I $\lambda6678$ lines in Figs.~\ref{profile_Ha}, \ref{prof_Ha2018}, and \ref{prof_HeI}, respectively. Limited by the relatively lower spectral resolution, we merely discuss the changes in line profiles qualitatively. 
The results shown in Figs. 5 and 6 of \citet{1998A&A...336..251N} exhibit similarities but with higher resolution and double-peaked profiles. We believe that most of our spectra are only single-peaked because of the low resolution of our observations. 

It can be seen from Fig.~\ref{profile_Ha} that the H$\alpha$ emission lines are presented a single peak most of the time; sometimes the peaks of the H$\alpha$ emission line are redshifted or blueshifted.
In Fig.~\ref{prof_Ha2018}, the H$\alpha$ emission line on January 07, 2018, presents a blueshifted, symmetric single-peaked profile. The peak intensity of the H$\alpha$ line becomes stronger during the November 03, 2019, observation. The H$\alpha$ emission line on December 26, 2019, presents a redshifted, symmetric single-peaked profile, which is observed just after a normal X-ray outburst. Moreover, the wings of the H$\alpha$ line on December 26, 2019, also become much broader than before. In 2020, the peak intensities of the H$\alpha$ lines become weaker with symmetric single-peaked profiles. The H$\alpha$ emission line on October 10, 2021, presents a double-peaked profile with V/R > 1 with a lower peak intensity than that in 2020.

As shown in Fig.~\ref{prof_HeI}, the He I $\lambda6678$ lines do not present emission features during our 2018 observations. The He I $\lambda6678$ line presents a double-peaked profile with V/R \textgreater 1 during our observation on November 03, 2019. Unlike the H$\alpha$ line, the He I $\lambda6678$ line reaches the local maximal intensity during our observation on October 10, 2020, one month ahead of the start of the giant X-ray outburst in 2020. We also note that the peak intensities of the He I $\lambda6678$ lines decrease during the onset of the 2020 giant outburst and continue to decrease during and after it. The He I $\lambda6678$ emission line on October 10, 2021, presents a double-peaked profile with V/R < 1, which is different from the H$\alpha$ lines.

\begin{figure*}[!htb]
\begin{minipage}{0.7\textwidth}
\includegraphics[bb=0 300 1000 2700,width=0.7\textwidth]{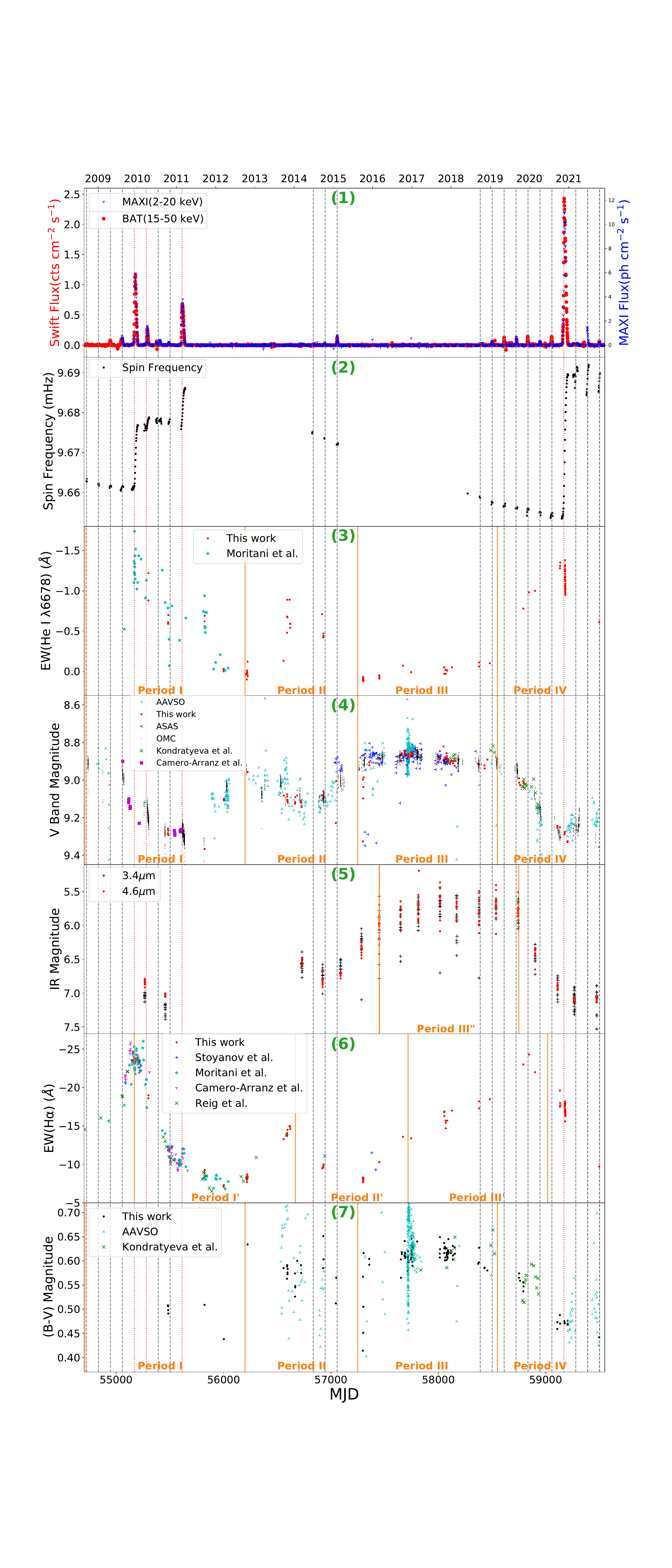}
\end{minipage}%
\begin{minipage}{0.3\textwidth}
\caption{Long-term variations of 1A 0535+262 from 2010 to 2021.
  First panel: X-ray band light curves. The data from \emph{Swift}-BAT and \emph{MAXI} are labeled with red points and blue downward-pointing triangles, respectively. January 1 of each year is also marked at the top of the first panel.
  Second panel: Barycentered and orbit-corrected spin-frequency history measured with \emph{Fermi}-GBM. 
  Third panel: Equivalent widths of He I $\lambda6678$ lines. The data from this work are labeled with red points. The data from \citet{2013PASJ...65...83M} are labeled with cyan stars.
  Fourth panel: Long-term light curves of optical {\it V} band. The data from \emph{AAVSO},  \emph{ASAS-SN}, \emph{INTEGRAL}-OMC, and this work are separately labeled with cyan upward-pointing triangles, blue leftward-pointing triangles, black pixels, and red points, respectively. In addition, the optical photometric data from \citet{2020Ap.....63..367K, 2012ApJ...754...20C} are labeled with green crosses and purple crosses, respectively.
  Fifth panel: Long-term light curves of \emph{NEOWISE} infrared magnitudes at 3.4 $\mu$m and 4.6 $\mu$m. These are labeled with black crosses and red points, respectively. 
  Sixth panel: Equivalent widths of H$\alpha$ lines. Our data are labeled with red points. The data from \citet{2020Ap.....63..367K, 2016ATel.8633....1S, 2013PASJ...65...83M, 2012ApJ...754...20C} are labeled with green crosses, blue crosses, cyan stars, and purple downward-pointing triangles, respectively. 
  Seventh panel: Evolution of $(B-V)$ color index. The data from \citet{2020Ap.....63..367K}, \emph{AAVSO,} and this work are labeled with green crosses, cyan upward-pointing triangles, and red points, respectively.
  The vertical dash-dotted lines indicate the times of the NS periastron passages at the ephemeris of MJD 53613.0+111.1E \citep{1996ApJ...459..288F}; of those the gray lines are close to normal X-ray outbursts, and the red lines are close to giant X-ray outbursts. The vertical solid lines indicate the division of observations into four distinct periods.
  \label{EW}}
\end{minipage}
\end{figure*}

\subsection{Long-term variations of He I $\lambda6678$ lines}

Periods I and II last from MJD 54700 to MJD 57250 (period I and period II are divided by MJD 56200.). On MJD 55820 and MJD 56600, there are two peaks in EW(He I $\lambda6678$), which are not accompanied by any X-ray outbursts. Apart from these two peaks, from MJD 55000 to MJD 55850, the He I $\lambda6678$ lines present emission lines. From 55900 to MJD 56200, the He I $\lambda6678$ lines cannot be distinguished from noise. From MJD 56590 to MJD 56930, the He I $\lambda6678$ lines also present emission lines, whose intensities are smaller than that during MJD 55000 to MJD 55850. 
During period III (MJD 57250--MJD 58550), the He I $\lambda6678$ lines cannot be distinguished from noise either. 
During period IV (MJD 58550--), the He I $\lambda6678$ lines present emission lines whose intensities are equal to that during MJD 55000 to MJD 55850. After the giant X-ray outburst in 2020, EW(He I $\lambda6678$) decreases to a lower value of $-$0.6 \AA \ in October 2021.
  

\subsection{Long-term variations of light curve of the optical {\it V} band}

As shown in the fourth panel of Fig.~\ref{EW}, from MJD 54700 to MJD 57250, there are two deep valleys and one peak in the {\it V}-band magnitudes. From MJD 54700 to MJD 55800, several normal X-ray outbursts and two giant X-ray outbursts take place in a phase of decreased optical brightness.  During MJD 55800 to MJD 56200, the {\it V}-band apparent magnitudes of 1A 0535+262 increase, reaching a local brightest magnitude with $M_{\it V} \backsim 8.95^{m}$. From MJD 56200 to MJD 57250, the brightness of the system first decreases and then increases to a magnitude of $M_{V} \backsim 8.9^{m}$ at the end of period II. Period II is shorter in time duration and smaller in magnitude changes than period I.
From MJD 57250 to MJD 58550, there is a plateau in the {\it V}-band magnitudes with $M_{V} \backsim 8.9^{m}$. 
From MJD 58550 to MJD 59170, there is phase of decrease in the {\it V}-band magnitudes until the giant X-ray outburst in 2020. After the giant X-ray outburst, the {\it V}-band magnitudes start to increase. 
It is likely that the {\it V}-band magnitudes and EW(He I $\lambda6678$) present an anti-correlation.

\subsection{Long-term variations of the light curves of infrared 3.4 $\mu$m and 4.6 $\mu$m}

Based on the data from \emph{NEOWISE}, only period III" (MJD 57450--58750) is unbroken. 
Following the {\it V} band, the magnitudes of infrared 3.4$\mu$m and 4.6$\mu$m present a similar trend from 2014 to 2020, but this lags behind that seen in the {\it V} band by $\backsim$200 days. The solid orange lines in the fifth panel of Fig.~\ref{EW} are 200 days behind that in the fourth panel.

\subsection{Long-term variations of equivalent widths of H$\alpha$ lines}

As shown in the sixth panel of Fig.~\ref{EW}, from MJD 55168 to MJD 57718, there are also two deep valleys and one peak in EW(H$\alpha$).
During period I' (MJD 55168--56668), the first valley of EW(H$\alpha$) is located about 500 days later than that of the {\it V}-band magnitudes. Near MJD 56620, EW(H$\alpha$) reaches a local maximum of $-$15 \AA.
During period II' (MJD 56668--57718), the second valley of EW(H$\alpha$) is also located about 500 days later than the second valley in {\it V}-band magnitude. During period II', several normal X-ray outbursts take place after EW(H$\alpha$) reaches the local maximum at the end of period I'. 
Unlike the {\it V} band, in Period III' (MJD 57718--59018), EW(H$\alpha$) continues to increase instead of presenting an apparent plateau phase. A peak of EW(H$\alpha$) can be seen on about MJD 58845, reaching a peak value of $-$24 \AA. After MJD 58845, EW(H$\alpha$) presents a phase of decrease until October 2021. The solid orange line in the sixth panel of Fig.~\ref{EW} is 468 days behind that in the fourth panel.

\subsection{Long-term variations of the $(B-V)$ color index}

Because of the low cadence and the large dispersion of the data, the $(B-V)$ color index does not present detailed changes in period I and II except an unclear trend of  increasing magnitude.
In period III and IV, the $(B-V)$ color index presents a similar trend to that of the {\it V} band. 
Namely, from MJD 57250 to MJD 58550, there is a plateau in the $(B-V)$ color index; from MJD 58550 to MJD 59170, there is a phase of  decrease in the $(B-V)$ color index until the giant X-ray outburst in 2020. 
After the giant X-ray outburst, the $(B-V)$ color index seems to increase, but this increase is not as clear as that of the {\it V} band. 

\subsection{Fast-photometry frequency analysis}

\begin{figure}
   \centering
   \includegraphics[width=\hsize]{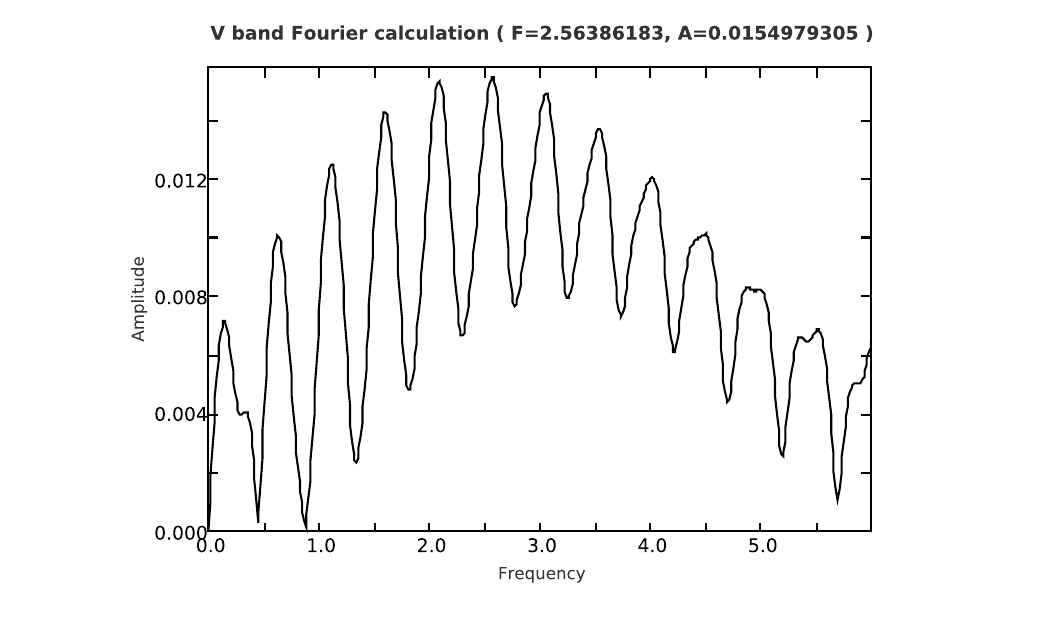}
     \caption{Multiperiodicity frequency of {\it V} band computed using Period04.
              }
         \label{V_cut} 
   \end{figure}

\begin{table}
\centering
\caption{\,Number of photometric measurements in B,V, and R bands from November 20, 2020, to November 24, 2020.}
\label{phot_frequency}
\begin{tabular}{ccccc}
\hline\hline
Date & Time of duration & B & V & R \\
& (h) &  &  &  \\
\hline
Nov. 20, 2020 & 1.13 & 52 & 52 & 52  \\   
Nov. 22, 2020 & 5.91 & 234 & 234 & 234 \\  
Nov. 24, 2020 & 2.73 & 79 & 79 & 79 \\
\hline
\end{tabular}
\end{table}
We used the code Period04 \citep{2005CoAst.146...53L} to analyze the frequency of the differential photometric data provided by the  Xinglong 80 cm telescope during the time period from November 20, 2020, to November 24, 2020, through standard Fourier analysis techniques. The number of measurements and their distribution over the observing nights are given in Table~\ref{phot_frequency}. 
We detected multiperiodicity frequencies in the light curve between 0 and 6 c d$^{-1}$. The main frequency of the {\it B, V,} and {\it R} bands are about 2.5--2.6 c d$^{-1}$, and the amplitudes are about 0.013--0.015 mag. The {\it V}-band result is presented in Fig.~\ref{V_cut}.

\section{Discussion} \label{sec:Discu}



\subsection{Correlation between observational phenomenons and mass transfer}


It is believed that the entire circumstellar disk contributes to the optically thick H$\alpha$ emission line in a Be star \citep{1992ApJS...81..335S}, whereas only the inner part of the circumstellar disk contributes mainly to the continuum flux, including the {\it V} band and near-infrared (NIR) bands \citep{2011IAUS..272..325C}. Due to the higher ionization potential energy, the emitting region of He I $\lambda6678$ must be limited to the inner part of the circumstellar disk, which is also smaller than the  extent of the nearby continuum \citep{1998A&A...332..268S}. In general, the emitting regions of He I $\lambda6678$, {\it V} band, and IR bands are from the inner disk to the outer disk in that order; whereas the emitting region of H$\alpha$ is the whole circumstellar disk, which is much bigger than the IR-band emitting region. Therefore, the successive changes in their intensity is likely caused by the viscous diffusion of matter on the circumstellar disk.

Let us discuss the correlations between the observational phenomena and mass transfer inside the disk of our target Be star, and between the disk and the NS in detail from 2011 to 2021. 
Just 200 days after the giant X-ray outburst in 2011, on MJD 55820, EW(He I $\lambda6678$) is at its peak value; the {\it V}-band magnitude is at its valley value; and there is no corresponding change in EW(H$\alpha$) at the same time. There is another peak of EW(He I $\lambda6678$) on MJD 56600. At this moment, EW(He I $\lambda6678$) and EW(H$\alpha$) are all at peak values. These two peaks of EW(He I $\lambda6678$) may be caused by mass ejection from the Be star to its surrounding disk. 
From MJD 56800 onwards, EW(He I $\lambda6678$) and EW(H$\alpha$) slowly decrease in intensity, and the {\it V}-band magnitude slowly increases. At this time, the direction of material flow is from the inner part to the outer part of the Be star disk and then to the NS, causing several normal X-ray outbursts.
From MJD 57250 onwards, it is likely that the matter ejection from the Be star decreases and then ceases, which is why EW(He I $\lambda6678$) and the {\it V}-band magnitude remain unchanged for a long time. However, due to the slow diffusion of the last quantities of matter ejected from the Be star disk, the infrared and EW (H$\alpha$) still increase for hundreds of days before gradually stabilizing.
From period IV onwards, the {\it V}-band magnitude decreases slowly, whereafter the infrared decreases. Meanwhile, EW(H$\alpha$) reaches its peak near MJD 58845. This observed phenomenon can be explained by the matter flowing from the inner part to the outer part of the Be star disk. EW(He I $\lambda6678$) should decrease in this condition. But on the contrary, EW(He I $\lambda6678$) increases rapidly before MJD 58845. 
Because the separation between the peaks of the He I $\lambda6678$ emission line we observed is not as large as that observed by \citet{2007A&A...475..651G}, we cannot be sure whether the He I $\lambda6678$ line emissions come from the accretion disk around the NS or not. However, since the peak of EW(He I $\lambda6678$) always occurs near the X-ray outbursts, it is very likely that part of the He I $\lambda6678$ line emission comes from the accretion disk of the NS \citep[see e.g.,][]{1990AcA....40...95G}. This also explains why the He I $\lambda6678$ and H$\alpha$ lines behave differently \citep[see e.g.,][]{1998MNRAS.294..165C}; we can also see an inconsistency in the H$\alpha$ and He I lines in 2021, which is discussed in Sect.~\ref{sec:Discu_V/R}. It is worth mentioning that the existence of the temporary accretion disk of the source during the X-ray outbursts has been confirmed \citep{1996ApJ...459..288F}, and that the He II emission line is a clear indicator of the presence of an accretion disk around compact stars in X-ray binaries \citep{2007A&A...475..651G}; a clear example being the He II $\lambda4686$ emission line that \citet{1978ApJ...223..530H} observed in JD 2443163 and JD 2443171.


In the long-term multiwavelength observations of 1A 0535+262, the variations of the He I $\lambda6678$ line are generally observed first. However, as part of the He I $\lambda6678$ line emission is likely from the accretion disk around the NS, He I $\lambda6678$ line emission from the accretion disk will contaminate the trend we see. Therefore, we generally base the conclusions outlined below on data for the {\it V} band, infrared, and H$\alpha$ lines. Following the {\it V} band, the magnitudes of infrared 3.4$\mu$m/4.6$\mu$m present similar variations to those of the {\it V} band but with a lag of $\backsim$200 days. Similarly, the variations of EW(H$\alpha$) show a similar trend to the {\it V}-band magnitudes, but with a longer time lag of $\backsim$500 days. We posit that it takes several hundred days for mass to transfer from the inner part to the outer part of the Be star disk.

There is one more thing to say:
On approximately MJD 56620, EW(H$\alpha$) exceeded $-$13 \AA; from MJD 56810 onwards, three normal X-ray outbursts took place. From MJD 57800 onwards, EW(H$\alpha$) stands above $-$13 \AA \ for most of the time, while several normal X-ray outbursts and one giant X-ray outburst take place. Based on the computation of \citet{2012ASPC..464..285M}, EW(H$\alpha$) estimated from the Be disk with the Roche lobe radius at periastron (5.6 Be star radii) is $-$13 \AA. We therefore have reason to believe that mass will accrete onto the NS when EW(H$\alpha$) exceeds $-$13 \AA. 



\subsection{Variations of equivalent width of the H$\alpha$ and He I $\lambda6678$ lines close to the giant X-ray outburst in 2020}

The equilibrium period $P_{\rm eq} \propto \dot{M}^{-3/7}$ \citep{2011Ap&SS.332....1R}.
From MJD 58380 to MJD 58730, EW(H$\alpha$) is just above $-$ 13\AA. The NS does not accrete a significant amount of matter, and so $\dot{M}$ and $P_{\rm eq}$ do not change significantly. Therefore, only several normal X-ray outbursts happen at that time. This is similar to the situation before the giant X-ray outburst in 2009.
The peak of EW(H$\alpha$) in 2019 happened one year before the giant X-ray outburst in 2020. This phenomenon also occurred in 1994 and 2005, but it did not happen in 2009, which can be seen in Figs. 3 and 4 of \citet{2012ApJ...744...37Y}. The decrease in the intensity of EW(H$\alpha$) from 2019 to 2020 is probably caused by mass transfer from the Be star disk to the NS. Therefore, we posit that giant X-ray outbursts may require an  accumulation of mass two or three hundred days before them. Meanwhile, the optical {\it V}-band and IR-band light curves of 1A 0535+262 are in a fading phase, which indicates that a mass-ejection event is happening at that time. This is similar to the situation in 1994, 2005, and 2009 \citep{1999MNRAS.302..167C, 2012ApJ...744...37Y, 2012ApJ...754...20C}. \citet{2004MNRAS.350.1457H} interpreted the relationship between dimming of the {\it V}-band and IR bands and the onset of the X-ray activity in 1A 0535+262 as a change of the resonant truncation of the Be star disk from a larger radius to a smaller one, which left material outside of the disk and available to be accreted by the NS. This must be what happened in these two or three hundred days.


\subsection{Variation in $(B-V)$ color index and inclination of the Be star disk} \label{sec:B-V}

\begin{figure}
   \centering
   \includegraphics[bb=0 0 600 400,width=\hsize]{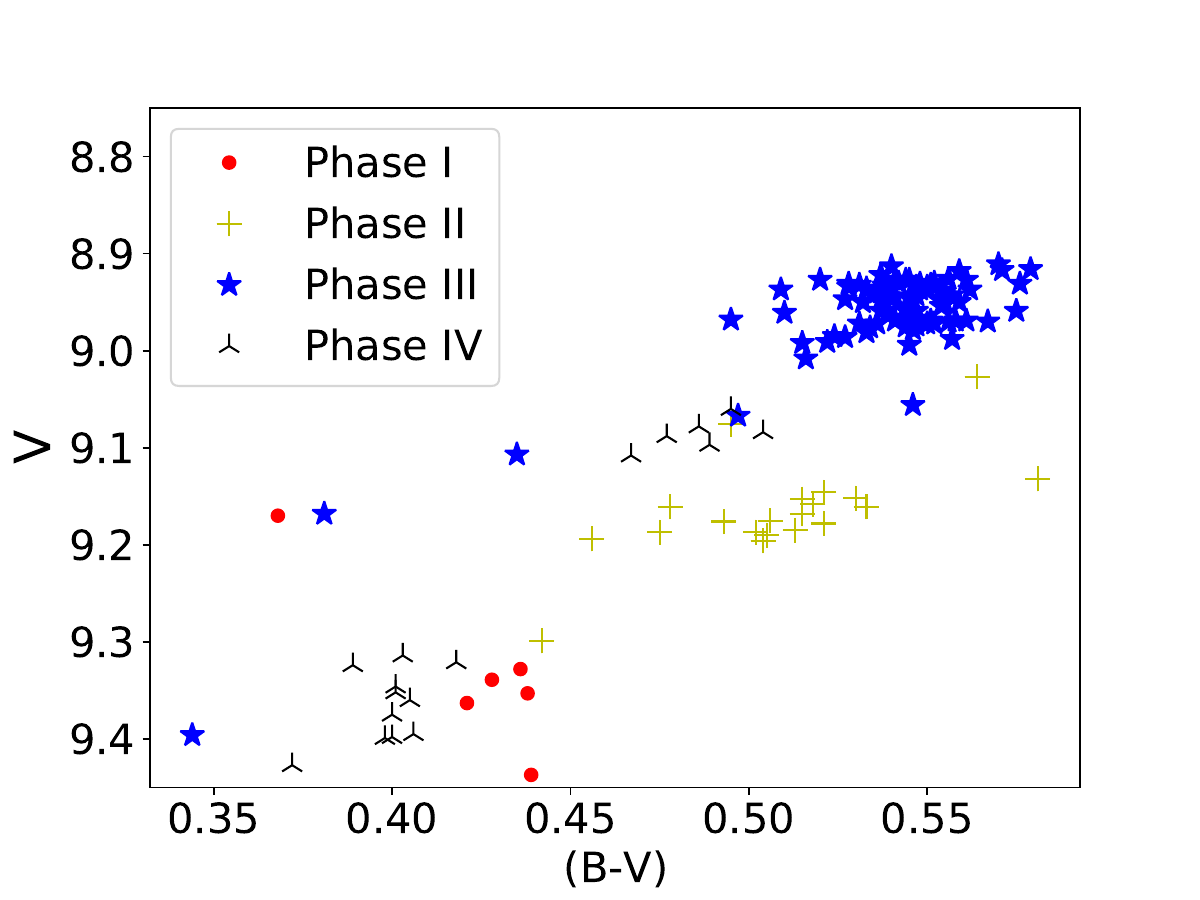}
      \caption{Variation of $(B-V)$ color index versus $V$-band magnitude.
              }
         \label{B-V_V}
   \end{figure}

Figure~\ref{B-V_V} shows the $(B-V)$ color index as a function of $V$ magnitude. We mark the different variability periods defined above with different colors. It has been noted that this kind of plot can be used to constrain the inclination angle of the system \citep{1983HvaOB...7...55H}.
Systems that show a positive correlation ---that is, as the disk forms (or equivalently, as EW(H$\alpha$) increases), the optical intensity increases and the emission becomes redder (i.e., $(B-V)$ increases)--- are thought to be seen at small or moderate inclination angles, while systems that show a negative correlation, in which the optical intensity decreases even though the disk is growing (EW(H$\alpha$) and $(B-V)$ increase), are associated with large inclination angles.

\citet{1983HvaOB...7...55H} introduced the concept of a pseudophotosphere to explain this effect. At large inclination angles (for equator-on stars), the inner parts of the Be envelope partly block the stellar photosphere, and thus the optical brightness decreases. Meanwhile, the overall emission becomes redder because the contribution of the disk increases. At small or moderate inclination, as the disk grows, an overall (star plus disk) increase in brightness is expected.

Figure~\ref{B-V_V} suggests that 1A 0535+262 is viewed at small or intermediate angles. This is consistent with the previous reported orbital inclination of \emph{i} = $37^{\circ} \pm 2^{\circ}$ \citep{2007A&A...475..651G}. 
At small, moderate, and large inclinations, the H$\alpha$ lines typically show a single-peak profile, a double-peak profile, and a double-peak with a central depression, respectively \citep{2013A&ARv..21...69R}.
1A 0535+262 shows single-peaked profiles in most cases (Fig.~\ref{profile_Ha}), which also suggests that the Be disk is viewed at a small inclination. 

In Fig.~\ref{B-V_V}, the data points of periods I and IV are in the lower left, and the data points of periods II and III are in the middle and upper right, respectively. A similar phenomenon of $(B-V)$ color index versus $V$-band magnitude was reported by \citet{1999MNRAS.302..167C}.
It is possible that periods I$\to$II$\to$III$\to$IV constitute a cycle. The data points in the lower left corner are observed when the X-ray outbursts occur. These points are bluer in color and dimmer in brightness, indicating that the radiation of the circumstellar disk becomes weaker and the proportion of the total radiation coming from the circumstellar disk reduces. The upper right points are observed in X-ray quiescence during period III; these points are redder and brighter, indicating that the radiation of the circumstellar disk becomes stronger at this time and that it makes up a greater proportion of the total radiation. 
It is worth mentioning that a 1400--1500 d cyclical behavior of 1A 0535+262 in $m_{K}$ versus EW(H$\alpha$) and EW(He I $\lambda6678$) was already mentioned by \citet{2004MNRAS.350.1457H}, and these authors believed it originated from Be disk precession period.



\subsection{Variability of the H$\alpha$ line profile and V/R ratio} \label{sec:Discu_V/R}

\begin{figure}
   \centering
   \includegraphics[bb=0 0 650 400,width=\hsize]{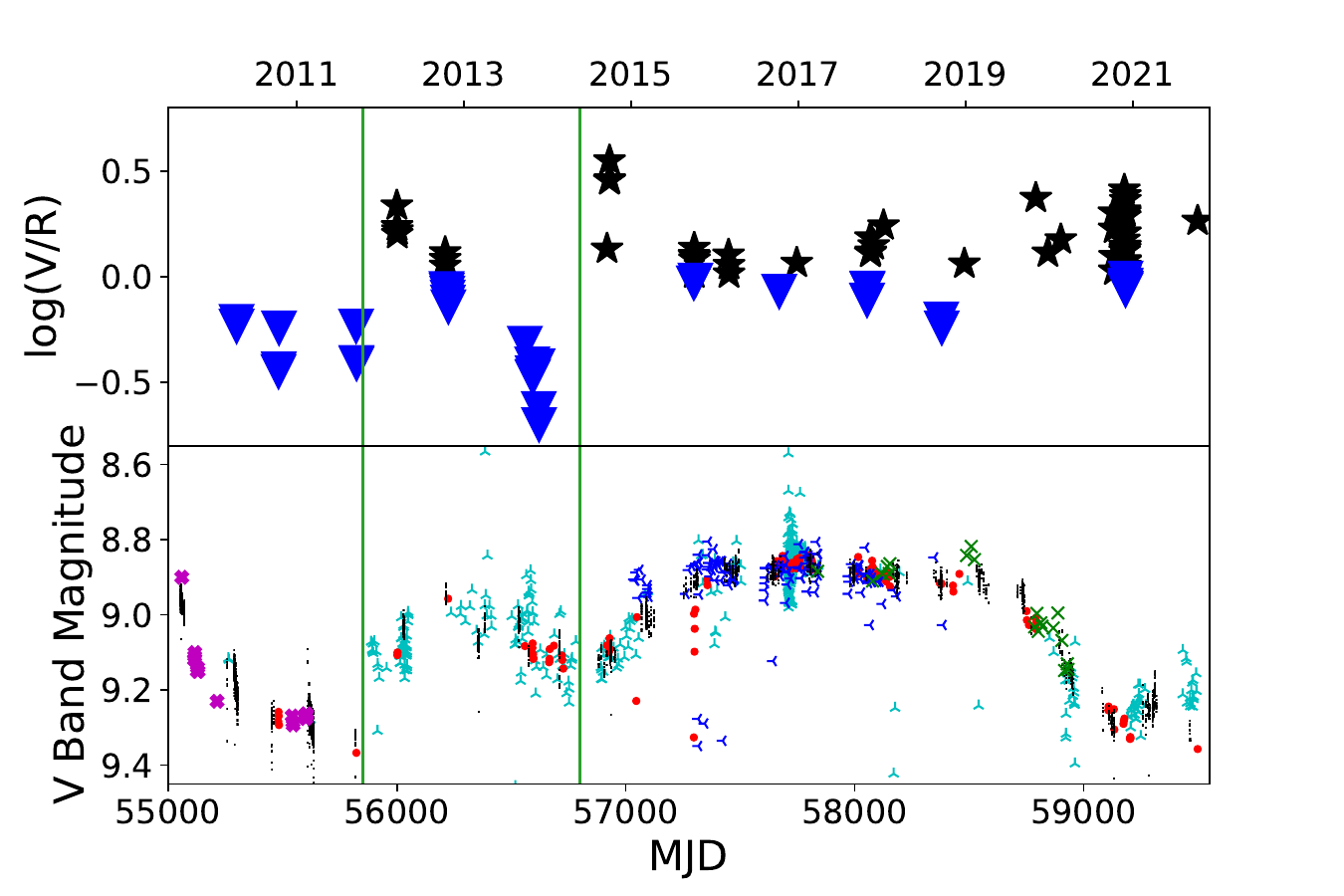}
      \caption{Long-term variations of 1A 0535+262 in $\log(V/R)$ and optical {\it V} band.
      Top panel: Evolution of $\log(V/R)$. 
       We use different symbols to represent the different profiles of spectral lines: stars represent the spectral profiles with V \textgreater\ R, triangles represent the spectral profiles with V \textless\ R.
       Bottom panel: Long-term light curves of optical {\it V} band. We omit the legend for conciseness; it is the same as the fourth panel of Fig.~\ref{EW}. The vertical green lines mark the minimum brightness in the photometric $V$ band at the V = R phase before the start of the V \textgreater\ R phase at $\sim$\,MJD 55850 and 56800, respectively.
              }
         \label{V_R}
   \end{figure}

The V/R variability is defined as the ratio V/R = (I(V) $-$ I$_\mathrm{c}$) / (I(R) $-$ I$_\mathrm{c}$) (where I(V), I(R), and I$_\mathrm{c}$ are the intensities of the violet peak, red peak, and continuum, respectively.), which is the ratio between the violet and red peak intensity of optically thin double emission lines referring to the continuum level \citep{1997A&A...326.1167M}, although the definition has been extended to include optically thick asymmetric emission profiles with subordinate peaks or shoulders \citep{1995A&A...302..751H,1995A&A...300..163H}.  In many Be stars, these variations are quasi-periodic if the star is monitored over a long enough period of time \citep{1997A&A...318..548O}. 

We also measured the separation of the violet and red peaks by fitting two Gaussian functions to the spectral line profiles. When the disk velocity is assumed to be Keplerian, the peak separation gives a measure of the velocity field.
There is no obvious trend in the peak separation between different spectral line profiles, which is mainly distributed around 150--250 km s$^{-1}$ (Table \ref{V_R_table}).

The V/R variability has been associated with density perturbations in the disk \citep{1995A&A...300..163H}. When this density perturbation moves around inside the disk, the profile changes. 
In Fig.~\ref{V_R}, we observe a red-dominated profile (V \textgreater\ R) in 2010--2011 that turns into a blue-dominated profile (V \textless\ R) in March 2012. In October 2012, we observe an almost single-peak profile (V$\sim$R). The spectra in 2013 return to blue-dominated profiles (V \textgreater\ R). 
These observations may therefore have covered an entire V/R cycle. The V/R quasi-period would be about two years, which is normal for Be/X-ray binaries \citep{1997A&A...326.1167M}.
The V/R quasi-period for 1A 0535+262 was about one year during 1994--1995 \citep{1998MNRAS.294..165C}. This is not absolutely consistent with the behavior of 1A 0535+262 from 2011 to 2015, when V/R shows a  slower change than that observed by \citet{1998MNRAS.294..165C}.
Therefore, we can perhaps say that the V/R quasi-period switched to a longer period during 2011--2015.

What is more interesting is that the V/R quasi-period disappears after 2015. From 2015 onwards, V/R remains a plateau until 2018. There is no obvious V/R quasi-period during 2018--2021.
The V/R ratio of the H$\alpha$ and He I lines in 2021 is different (see Figs.~\ref{prof_Ha2018} and \ref{prof_HeI}). Indeed, the $\log(V/R)$ of the H$\alpha$ lines in 2021 is positive, whereas the $\log(V/R)$ of the He I $\lambda6678$ lines in 2021 is negative. As our observation in 2021 was made near the normal X-ray outburst, it is very likely that part of the He I $\lambda6678$ line emission comes from the accretion disk of the NS. 

In principle, whether the motion of the perturbation occurs in the same sense (prograde rotation) or in the opposite sense (retrograde rotation) to the stellar rotation can be determined from the observations. 
\citet{1994A&A...288..558T} realized that a prograde rotation implies (i) a V \textgreater\ R phase, (ii) a shell absorption profile, (iii) a V \textless\ R phase, and (iv) a weak central absorption profile will appear, in that order. A retrograde rotation would give rise to the reversed sequence: (iv)$\to$(iii)$\to$(ii)$\to$(i). Because of the small disk inclination (see \ref{sec:B-V}), we cannot distinguish between a prograde and a retrograde rotation in the characteristic line shapes. However, these characteristic line shapes can translate into noticeable photometric variations.
According to \citet{1997A&A...326.1167M}, we can expect a minimum brightness when V = R prior to the V \textless\ R (V \textgreater\ R) phase if the motion is prograde (retrograde).
In 1A 0535+262, the minimum brightness in the photometric $V$ band (the vertical green lines in Fig.~\ref{V_R}) occurred during the V = R phase before the V \textgreater\ R phase began, on $\sim$\,MJD 55850 and 56800, confirming the retrograde nature of the precession inside the disk.

\section{Conclusions} \label{sec:Concl}

1A 0535+262 was X-ray-active from the second half of 2018 to 2021, including several normal outbursts and a giant outburst. From 2010 to 2021, 1A 0535+262 was observed by spectroscopy and photometry in order to study the Be disk structure and mass transfer between the Be star circumstellar disk and the NS over a long timescale. In particular, our data cover the observations one month and one year before the giant X-ray outburst in 2020 with medium-dispersion spectrographs in the wavelength range of 4000--8700 \AA, which includes H$\alpha$ and He I $\lambda6678$ lines.

The emission regions of {\it V} band, IR bands, and H$\alpha$ on the circumstellar disk come from the inner part to the outer part, respectively, and so the intensity change from the inner to the outer disk is likely caused by the viscous diffusion of material on the Be star circumstellar disk. It usually takes several hundred days for mass to transfer from the inner part to the outer part of the circumstellar disk and to be accreted by the NS.
Once EW(H$\alpha$) exceeds $-$13 \AA, the process of mass accretion begins. For the giant X-ray outburst in 2020, it takes about one year for material accumulation. 

As (i) EW(He I $\lambda6678$) increases rapidly and abnormally from MJD 58790 to MJD 59180, (ii) $\log(V/R)$ of the H$\alpha$ and He I $\lambda6678$ lines in 2021 is different, and (iii) our observation in 2021 was taken near the normal X-ray outburst, it is very likely that part of the He I $\lambda6678$ line emission comes from the accretion disk around the NS.


\begin{acknowledgements}   
      We acknowledge the support of the staff of the Xinglong 2.16m telescope, the Xinglong 80 cm telescope and the Xinglong 60 cm telescope. This work was partially supported by the Open Project Program of the CAS Key Laboratory of Optical Astronomy, National Astronomical Observatories, Chinese Academy of Sciences.
      
      We acknowledge the support of the staff of the Lijiang 2.4m telescope. Funding for the telescope has been provided by CAS and the People's Government of Yunnan Province.
      
      We acknowledge the spectrum datum from \citet{2020Ap.....63..367K}, which was observed on February 20, 2020.
      
      We acknowledge with thanks the variable star observations from the \emph{AAVSO} International Database contributed by observers worldwide and used in this research.
   
      \emph{Swift}-BAT transient monitor results provided by the \emph{Swift}-BAT team. \emph{Fermi}-GBM results provided by the Fermi Science Support Center.
      
      This publication makes use of data products from \emph{NEOWISE}, which is a project of the Jet Propulsion Laboratory/California Institute of Technology, funded by the Planetary Science Division of the National Aeronautics and Space Administration.
      
      This work is supported by the National Natural Science Foundation of China (Grants No. U2031205, 11733009).

      We thank the anonymous referee for her/his useful comments.
\end{acknowledgements}



\bibliographystyle{aa} 
\bibliography{1A0535} 

\clearpage
\onecolumn

\begin{appendix}

\section{Tables of spectroscopic observations}

\begin{longtable}{cclccc}
\caption{Summary of spectroscopic observations of 1A 0535+262.\label{table1}}\\
\hline
\hline
Date & MJD & Telescope/ & Wavelength range & EW(H$\alpha$) & EW(He I $\lambda6678$) \\
(DD-MM-YYYY) & & Instrument & (\AA) & (\AA) & (\AA) \\
\hline
\endfirsthead
\caption{Continued.} \\
\hline
\hline
Date & MJD & Telescope/ & Wavelength range & EW(H$\alpha$) & EW(He I $\lambda6678$) \\
(DD-MM-YYYY) & & Instrument & (\AA) & (\AA) & (\AA) \\
\hline
\endhead
\hline
\endfoot
\hline
\endlastfoot
 13-04-2010 & 55299.5259 & 2.16 m/BFOSC & 5950-8200 & -18.6 $\pm$ 0.1 & -1.22 $\pm$ 0.20\\
 13-04-2010 & 55299.5340 & 2.16 m/BFOSC & 5950-8200 & -19.0 $\pm$ 0.1 & -0.88 $\pm$ 0.06\\
 13-10-2010 & 55482.8702 & 2.16 m/OMR   & 5500-6900 & -11.9 $\pm$ 0.1 & -0.59 $\pm$ 0.15\\
 14-10-2010 & 55483.7825 & 2.16 m/OMR   & 5500-6900 & -11.8 $\pm$ 0.1 & -0.61 $\pm$ 0.12\\
 15-10-2010 & 55484.8245 & 2.16 m/OMR   & 5500-6900 & -12.3 $\pm$ 0.1 & -0.70 $\pm$ 0.14\\
 17-09-2011 & 55821.8229 & 2.16 m/OMR   & 5500-6900 &  -9.2 $\pm$ 0.1 & -0.62 $\pm$ 0.09\\
 18-09-2011 & 55822.8008 & 2.16 m/OMR   & 5500-6900 &  -8.9 $\pm$ 0.2 & -0.56 $\pm$ 0.11\\
 19-09-2011 & 55823.8159 & 2.16 m/OMR   & 5500-6900 &  -9.3 $\pm$ 0.1 & -0.69 $\pm$ 0.16\\
 13-03-2012 & 55999.6244 & 2.4 m /YFOSC & 4970-9830 &  -7.1 $\pm$ 0.1 & -0.01 $\pm$ 0.06\\
 14-03-2012 & 56000.6172 & 2.4 m /YFOSC & 4970-9830 &  -7.2 $\pm$ 0.1 & -0.03 $\pm$ 0.04\\
 15-03-2012 & 56001.6194 & 2.4 m /YFOSC & 4970-9830 &  -7.3 $\pm$ 0.1 &  0.00 $\pm$ 0.03\\
 16-03-2012 & 56002.5698 & 2.4 m /YFOSC & 4970-9830 &  -7.3 $\pm$ 0.1 & -0.00 $\pm$ 0.04\\
 11-10-2012 & 56210.8138 & 2.16 m/OMR   & 5200-7800 &  -8.2 $\pm$ 0.1 &  0.04 $\pm$ 0.08\\
 11-10-2012 & 56210.8153 & 2.16 m/OMR   & 5200-7800 &  -8.3 $\pm$ 0.1 &  0.03 $\pm$ 0.08\\
 11-10-2012 & 56210.8170 & 2.16 m/OMR   & 5200-7800 &  -8.3 $\pm$ 0.1 &  0.04 $\pm$ 0.09\\
 16-10-2012 & 56216.8552 & 2.16 m/OMR   & 5500-6900 &  -7.7 $\pm$ 0.1 &  0.10 $\pm$ 0.09\\
 17-10-2012 & 56217.7852 & 2.16 m/OMR   & 5500-6900 &  -8.0 $\pm$ 0.1 & -0.01 $\pm$ 0.07\\
 19-10-2012 & 56219.8293 & 2.16 m/OMR   & 5500-6900 &  -8.7 $\pm$ 0.1 &  0.01 $\pm$ 0.12\\
 21-10-2012 & 56221.8768 & 2.16 m/OMR   & 5500-6900 &  -8.2 $\pm$ 0.1 &  0.03 $\pm$ 0.09\\
 22-10-2012 & 56222.8718 & 2.16 m/OMR   & 5500-6900 &  -8.4 $\pm$ 0.1 & -0.12 $\pm$ 0.16\\
 24-10-2012 & 56224.8166 & 2.16 m/OMR   & 5500-6900 &  -8.3 $\pm$ 0.1 &  0.06 $\pm$ 0.21\\
 23-09-2013 & 56558.7990 & 2.16 m/OMR   & 5500-6900 & -13.3 $\pm$ 0.1 & -0.13 $\pm$ 0.15\\
 25-10-2013 & 56590.7687 & 2.16 m/OMR   & 5500-6900 & -14.1 $\pm$ 0.1 & -0.48 $\pm$ 0.16\\
 26-10-2013 & 56591.8121 & 2.16 m/OMR   & 5500-6900 & -13.7 $\pm$ 0.1 & -0.68 $\pm$ 0.22\\
 27-10-2013 & 56592.7368 & 2.16 m/OMR   & 5500-6900 & -13.9 $\pm$ 0.1 & -0.89 $\pm$ 0.22\\
 29-10-2013 & 56594.8093 & 2.16 m/OMR   & 5500-6900 & -14.5 $\pm$ 0.1 & -0.67 $\pm$ 0.13\\
 18-11-2013 & 56614.6347 & 2.4 m /YFOSC & 4970-9830 & -15.0 $\pm$ 0.1 & -0.89 $\pm$ 0.19\\
 23-11-2013 & 56619.6388 & 2.4 m /YFOSC & 4970-9830 & -14.9 $\pm$ 0.1 & -0.54 $\pm$ 0.14\\
 24-11-2013 & 56620.6076 & 2.4 m /YFOSC & 4970-9830 & -14.6 $\pm$ 0.1 & -0.59 $\pm$ 0.19\\
 17-09-2014 & 56917.8102 & 2.16 m/OMR   & 5500-6900 &  -9.5 $\pm$ 0.1 & -0.71 $\pm$ 0.14\\
 27-09-2014 & 56927.8187 & 2.16 m/OMR   & 5500-6900 &  -9.6 $\pm$ 0.1 & -0.42 $\pm$ 0.16\\
 28-09-2014 & 56928.8711 & 2.16 m/OMR   & 5500-6900 &  -9.8 $\pm$ 0.1 & -0.44 $\pm$ 0.13\\
 29-09-2014 & 56929.8474 & 2.16 m/OMR   & 5500-6900 & -10.0 $\pm$ 0.1 & -0.47 $\pm$ 0.15\\
 01-10-2015 & 57296.8319 & 2.16 m/BFOSC & 5950-8200 &  -8.3 $\pm$ 0.1 &  0.11 $\pm$ 0.16\\
 02-10-2015 & 57297.8116 & 2.16 m/BFOSC & 3900-6700 &  -7.7 $\pm$ 0.1 &  0.07 $\pm$ 0.02\\
 03-10-2015 & 57298.8026 & 2.16 m/BFOSC & 5950-8200 &  -8.2 $\pm$ 0.1 &  0.07 $\pm$ 0.08\\
 03-10-2015 & 57298.8875 & 2.16 m/BFOSC & 5950-8200 &  -8.1 $\pm$ 0.1 &  0.09 $\pm$ 0.05\\
 04-10-2015 & 57299.8633 & 2.16 m/BFOSC & 5950-8200 &  -8.0 $\pm$ 0.1 &  0.12 $\pm$ 0.06\\
 05-10-2015 & 57300.7628 & 2.16 m/BFOSC & 5950-8200 &  -7.9 $\pm$ 0.1 &  0.09 $\pm$ 0.12\\
 08-10-2015 & 57303.8633 & 2.16 m/BFOSC & 5950-8200 &  -8.2 $\pm$ 0.1 &  0.11 $\pm$ 0.08\\
 01-03-2016 & 57448.6247 & 2.4 m /YFOSC & 4970-9830 & -10.3 $\pm$ 0.1 &  0.05 $\pm$ 0.09\\
 03-03-2016 & 57450.5706 & 2.4 m /YFOSC & 4970-9830 & -10.3 $\pm$ 0.1 &  0.07 $\pm$ 0.06\\
 03-03-2016 & 57450.6352 & 2.4 m /YFOSC & 4970-9830 & -10.3 $\pm$ 0.1 &  0.09 $\pm$ 0.09\\
 08-10-2016 & 57669.8328 & 2.16 m/OMR   & 5500-6900 & -13.6 $\pm$ 0.1 & -0.07 $\pm$ 0.06\\
 24-12-2016 & 57746.6254 & 2.4 m /YFOSC & 4970-9830 & -13.4 $\pm$ 0.1 &  0.01 $\pm$ 0.07\\
 27-10-2017 & 58053.7641 & 2.16 m/OMR   & 5500-6900 & -16.8 $\pm$ 0.1 & -0.05 $\pm$ 0.12\\
 28-10-2017 & 58054.8780 & 2.16 m/OMR   & 5500-6900 & -16.3 $\pm$ 0.1 & -0.05 $\pm$ 0.17\\
 29-10-2017 & 58055.7968 & 2.16 m/OMR   & 5500-6900 & -16.9 $\pm$ 0.1 &  0.03 $\pm$ 0.10\\
 10-11-2017 & 58067.8338 & 2.16 m/OMR   & 5500-6900 & -15.7 $\pm$ 0.1 & -0.02 $\pm$ 0.20\\
 11-11-2017 & 58068.7906 & 2.16 m/OMR   & 5500-6900 & -15.4 $\pm$ 0.1 & -0.02 $\pm$ 0.19\\
 12-11-2017 & 58069.7903 & 2.16 m/OMR   & 5500-6900 & -14.7 $\pm$ 0.1 &  0.03 $\pm$ 0.17\\
 25-11-2017 & 58082.6411 & 2.4 m /YFOSC & 4970-9830 & -15.7 $\pm$ 0.1 & -0.01 $\pm$ 0.04\\
 07-01-2018 & 58125.6912 & 2.4 m /YFOSC & 4970-9830 & -17.0 $\pm$ 0.1 & -0.05 $\pm$ 0.04\\
 16-09-2018 & 58377.8165 & 2.16 m/OMR   & 5500-6900 & -18.2 $\pm$ 0.1 & -0.06 $\pm$ 0.10\\
 18-09-2018 & 58379.8221 & 2.16 m/OMR   & 5500-6900 & -17.3 $\pm$ 0.1 & -0.11 $\pm$ 0.10\\
 26-12-2018 & 58478.6664 & 2.4 m /YFOSC & 4970-9830 & -18.5 $\pm$ 0.1 & -0.10 $\pm$ 0.06\\
 03-11-2019 & 58790.8476 & 2.16 m/OMR   & 5500-6900 & -23.0 $\pm$ 0.2 & -0.78 $\pm$ 0.18\\
 26-12-2019 & 58843.7108 & 2.4 m /YFOSC & 4970-9830 & -24.3 $\pm$ 0.1 & -0.98 $\pm$ 0.13\\
 20-02-2020 & 58900.0960 & 0.7 m /AZT-8 & 6100-6800 & -22.0 $\pm$ 0.5 & -1.00 $\pm$ 0.28\\
 09-10-2020 & 59131.8682 & 2.16 m/BFOSC & 5950-8200 & -19.6 $\pm$ 0.1 & -1.28 $\pm$ 0.51\\
 10-10-2020 & 59132.6998 & 2.16 m/BFOSC & 5950-8200 & -17.5 $\pm$ 0.2 & -1.28 $\pm$ 0.20\\
 11-10-2020 & 59133.8590 & 2.16 m/BFOSC & 5950-8200 & -18.0 $\pm$ 0.1 & -1.31 $\pm$ 0.09\\
 12-10-2020 & 59134.7982 & 2.16 m/BFOSC & 5950-8200 & -17.7 $\pm$ 0.1 & -1.35 $\pm$ 0.34\\
 24-11-2020 & 59177.6292 & 2.16 m/BFOSC & 5950-8200 & -17.5 $\pm$ 0.1 & -1.14 $\pm$ 0.26\\
 24-11-2020 & 59177.6442 & 2.16 m/BFOSC & 5950-8200 & -17.8 $\pm$ 0.1 & -1.26 $\pm$ 0.07\\
 24-11-2020 & 59177.6591 & 2.16 m/BFOSC & 5950-8200 & -17.7 $\pm$ 0.1 & -1.27 $\pm$ 0.08\\
 25-11-2020 & 59178.7197 & 2.16 m/BFOSC & 5950-8200 & -17.2 $\pm$ 0.1 & -1.16 $\pm$ 0.15\\
 25-11-2020 & 59178.7408 & 2.16 m/BFOSC & 5950-8200 & -17.8 $\pm$ 0.1 & -1.30 $\pm$ 0.10\\
 26-11-2020 & 59179.7861 & 2.16 m/BFOSC & 5950-8200 & -15.6 $\pm$ 0.1 & -1.26 $\pm$ 0.07\\
 26-11-2020 & 59179.8077 & 2.16 m/BFOSC & 5950-8200 & -18.2 $\pm$ 0.1 & -1.24 $\pm$ 0.12\\
 26-11-2020 & 59179.8182 & 2.16 m/BFOSC & 5950-8200 & -18.0 $\pm$ 0.1 & -1.38 $\pm$ 0.11\\
 27-11-2020 & 59180.7689 & 2.16 m/BFOSC & 5950-8200 & -18.0 $\pm$ 0.1 & -1.21 $\pm$ 0.08\\
 27-11-2020 & 59180.7798 & 2.16 m/BFOSC & 5950-8200 & -18.0 $\pm$ 0.1 & -1.18 $\pm$ 0.14\\
 27-11-2020 & 59180.7868 & 2.16 m/BFOSC & 5950-8200 & -17.8 $\pm$ 0.1 & -1.32 $\pm$ 0.12\\
 27-11-2020 & 59180.7938 & 2.16 m/BFOSC & 5950-8200 & -17.8 $\pm$ 0.1 & -1.25 $\pm$ 0.14\\
 27-11-2020 & 59180.8008 & 2.16 m/BFOSC & 5950-8200 & -17.9 $\pm$ 0.1 & -1.27 $\pm$ 0.09\\
 28-11-2020 & 59181.8820 & 2.16 m/BFOSC & 5950-8200 & -16.8 $\pm$ 0.1 & -1.21 $\pm$ 0.13\\
 28-11-2020 & 59181.8893 & 2.16 m/BFOSC & 5950-8200 & -16.9 $\pm$ 0.1 & -1.17 $\pm$ 0.12\\
 28-11-2020 & 59181.8928 & 2.16 m/BFOSC & 5950-8200 & -16.7 $\pm$ 0.1 & -1.18 $\pm$ 0.13\\
 28-11-2020 & 59181.8963 & 2.16 m/BFOSC & 5950-8200 & -17.0 $\pm$ 0.2 & -1.06 $\pm$ 0.13\\
 28-11-2020 & 59181.8999 & 2.16 m/BFOSC & 5950-8200 & -17.0 $\pm$ 0.1 & -1.04 $\pm$ 0.17\\
 28-11-2020 & 59181.9034 & 2.16 m/BFOSC & 5950-8200 & -17.0 $\pm$ 0.1 & -1.07 $\pm$ 0.11\\
 28-11-2020 & 59181.9070 & 2.16 m/BFOSC & 5950-8200 & -16.7 $\pm$ 0.1 & -1.06 $\pm$ 0.16\\
 28-11-2020 & 59181.9105 & 2.16 m/BFOSC & 5950-8200 & -16.9 $\pm$ 0.1 & -1.06 $\pm$ 0.12\\
 28-11-2020 & 59181.9140 & 2.16 m/BFOSC & 5950-8200 & -17.2 $\pm$ 0.1 & -1.00 $\pm$ 0.11\\
 29-11-2020 & 59182.8790 & 2.16 m/BFOSC & 5950-8200 & -16.5 $\pm$ 0.1 & -1.09 $\pm$ 0.13\\
 29-11-2020 & 59182.8825 & 2.16 m/BFOSC & 5950-8200 & -16.4 $\pm$ 0.1 & -1.03 $\pm$ 0.16\\
 29-11-2020 & 59182.8860 & 2.16 m/BFOSC & 5950-8200 & -16.3 $\pm$ 0.2 & -1.02 $\pm$ 0.15\\
 29-11-2020 & 59182.8896 & 2.16 m/BFOSC & 5950-8200 & -16.4 $\pm$ 0.1 & -1.13 $\pm$ 0.16\\
 29-11-2020 & 59182.8931 & 2.16 m/BFOSC & 5950-8200 & -16.4 $\pm$ 0.1 & -0.95 $\pm$ 0.26\\
 29-11-2020 & 59182.8967 & 2.16 m/BFOSC & 5950-8200 & -16.4 $\pm$ 0.1 & -1.12 $\pm$ 0.25\\
 29-11-2020 & 59182.9002 & 2.16 m/BFOSC & 5950-8200 & -16.4 $\pm$ 0.1 & -1.01 $\pm$ 0.29\\
 29-11-2020 & 59182.9037 & 2.16 m/BFOSC & 5950-8200 & -16.5 $\pm$ 0.1 & -0.97 $\pm$ 0.28\\
 29-11-2020 & 59182.9073 & 2.16 m/BFOSC & 5950-8200 & -16.4 $\pm$ 0.1 & -0.98 $\pm$ 0.29\\
 29-11-2020 & 59182.9108 & 2.16 m/BFOSC & 5950-8200 & -16.4 $\pm$ 0.1 & -1.01 $\pm$ 0.26\\
 10-10-2021 & 59497.7769 & 2.16 m/OMR   & 6000-6850 &  -9.7 $\pm$ 0.1 & -0.61 $\pm$ 0.12\\
\end{longtable}

\begin{longtable}{cclccc}
\caption{V/R ratio and peak separation of 1A 0535+262.\label{V_R_table}}\\
\hline
\hline
Date & MJD & Telescope/ & Wavelength Range & $\log(V/R)$(H$\alpha$) & $\Delta$V(H$\alpha$)\\
(DD-MM-YYYY) & & Instrument & (\AA) & & (km s$^{-1}$)\\
\hline
\endfirsthead
\caption{Continued.} \\
\hline
\hline
Date & MJD & Telescope/ & Wavelength Range & $\log(V/R)$(H$\alpha$) & $\Delta$V(H$\alpha$)\\
(DD-MM-YYYY) & & Instrument & (\AA) & & (km s$^{-1}$)\\
\hline
\endhead
\hline
\endfoot
\hline
\endlastfoot
  13-04-2010 & 55299.5259 & 2.16 m/BFOSC & 5950-8200 & -0.213 & 189 \\
  13-04-2010 & 55299.5340 & 2.16 m/BFOSC & 5950-8200 & -0.235 & 192 \\
  13-10-2010 & 55482.8702 & 2.16 m/OMR   & 5500-6900 & -0.450 & 205 \\
  14-10-2010 & 55483.7825 & 2.16 m/OMR   & 5500-6900 & -0.437 & 202 \\
  15-10-2010 & 55484.8245 & 2.16 m/OMR   & 5500-6900 & -0.244 & 196 \\
  17-09-2011 & 55821.8229 & 2.16 m/OMR   & 5500-6900 & -0.235 & 173 \\
  18-09-2011 & 55822.8008 & 2.16 m/OMR   & 5500-6900 & -0.407 & 251 \\
  19-09-2011 & 55823.8159 & 2.16 m/OMR   & 5500-6900 & -0.412 & 261 \\
  13-03-2012 & 55999.6244 & 2.4 m/YFOSC  & 4970-9830 &  0.336 & 238 \\
  14-03-2012 & 56000.6172 & 2.4 m/YFOSC  & 4970-9830 &  0.238 & 220 \\
  15-03-2012 & 56001.6194 & 2.4 m/YFOSC  & 4970-9830 &  0.201 & 242 \\
  16-03-2012 & 56002.5698 & 2.4 m/YFOSC  & 4970-9830 &  0.218 & 263 \\
  11-10-2012 & 56210.8138 & 2.16 m/OMR   & 5500-6900 &  0.083 & 239 \\
  11-10-2012 & 56210.8153 & 2.16 m/OMR   & 5500-6900 &  0.051 & 245 \\
  11-10-2012 & 56210.8170 & 2.16 m/OMR   & 5500-6900 &  0.115 & 246 \\
  16-10-2012 & 56216.8552 & 2.16 m/OMR   & 5500-6900 & -0.057 & 228 \\
  17-10-2012 & 56217.7852 & 2.16 m/OMR   & 5500-6900 & -0.065 & 222 \\
  19-10-2012 & 56219.8293 & 2.16 m/OMR   & 5500-6900 & -0.079 & 223 \\
  21-10-2012 & 56221.8768 & 2.16 m/OMR   & 5500-6900 & -0.097 & 219 \\
  22-10-2012 & 56222.8718 & 2.16 m/OMR   & 5500-6900 & -0.121 & 214 \\
  24-10-2012 & 56224.8166 & 2.16 m/OMR   & 5500-6900 & -0.144 & 224 \\
  23-09-2013 & 56558.7990 & 2.16 m/OMR   & 5500-6900 & -0.316 & 166 \\
  25-10-2013 & 56590.7687 & 2.16 m/OMR   & 5500-6900 & -0.414 & 226 \\
  26-10-2013 & 56591.8121 & 2.16 m/OMR   & 5500-6900 & -0.474 & 241 \\
  27-10-2013 & 56592.7368 & 2.16 m/OMR   & 5500-6900 & -0.456 & 236 \\
  29-10-2013 & 56594.8093 & 2.16 m/OMR   & 5500-6900 & -0.473 & 222 \\
  18-11-2013 & 56614.6347 & 2.4 m/YFOSC  & 4970-9830 & -0.422 & 255 \\
  23-11-2013 & 56619.6388 & 2.4 m/YFOSC  & 4970-9830 & -0.625 & 266 \\
  24-11-2013 & 56620.6076 & 2.4 m/YFOSC  & 4970-9830 & -0.701 & 291 \\
  17-09-2014 & 56917.8102 & 2.16 m/OMR   & 5500-6900 &  0.132 & 183 \\
  27-09-2014 & 56927.8187 & 2.16 m/OMR   & 5500-6900 &  0.455 & 213 \\
  28-09-2014 & 56928.8711 & 2.16 m/OMR   & 5500-6900 &  0.549 & 227 \\
  29-09-2014 & 56929.8474 & 2.16 m/OMR   & 5500-6900 &  0.459 & 203 \\
  01-10-2015 & 57296.8319 & 2.16 m/BFOSC & 5950-8200 & -0.031 & 231 \\
  02-10-2015 & 57297.8116 & 2.16 m/BFOSC & 5950-8200 &  0.101 & 217 \\
  03-10-2015 & 57298.8026 & 2.16 m/BFOSC & 5950-8200 &  0.136 & 207 \\
  03-10-2015 & 57298.8875 & 2.16 m/BFOSC & 5950-8200 &  0.014 & 192 \\
  04-10-2015 & 57299.8633 & 2.16 m/BFOSC & 5950-8200 &  0.081 & 202 \\
  05-10-2015 & 57300.7628 & 2.16 m/BFOSC & 5950-8200 &  0.096 & 218 \\
  08-10-2015 & 57303.8633 & 2.16 m/BFOSC & 5950-8200 & -0.017 & 222 \\
  01-03-2016 & 57448.6247 & 2.4 m/YFOSC  & 4970-9830 &  0.104 & 206 \\
  03-03-2016 & 57450.5706 & 2.4 m/YFOSC  & 4970-9830 &  0.055 & 225 \\
  03-03-2016 & 57450.6352 & 2.4 m/YFOSC  & 4970-9830 &  0.014 & 245 \\
  08-10-2016 & 57669.8328 & 2.16 m/OMR   & 5500-6900 & -0.071 & 164 \\
  24-12-2016 & 57746.6254 & 2.4 m/YFOSC  & 4970-9830 &  0.065 & 270 \\
  27-10-2017 & 58053.7641 & 2.16 m/OMR   & 5500-6900 & -0.112 & 201 \\
  28-10-2017 & 58054.8780 & 2.16 m/OMR   & 5500-6900 & -0.055 & 200 \\
  29-10-2017 & 58055.7968 & 2.16 m/OMR   & 5500-6900 & -0.059 & 203 \\
  10-11-2017 & 58067.8338 & 2.16 m/OMR   & 5500-6900 &  0.112 & 173 \\
  11-11-2017 & 58068.7906 & 2.16 m/OMR   & 5500-6900 &  0.185 & 175 \\
  12-11-2017 & 58069.7903 & 2.16 m/OMR   & 5500-6900 &  0.118 & 192 \\
  25-11-2017 & 58082.6411 & 2.4 m/YFOSC  & 4970-9830 &  0.148 & 226 \\
  07-01-2018 & 58125.6912 & 2.4 m/YFOSC  & 4970-9830 &  0.242 & 198 \\
  16-09-2018 & 58377.8165 & 2.16 m/OMR   & 5500-6900 & -0.201 & 181 \\
  18-09-2018 & 58379.8221 & 2.16 m/OMR   & 5500-6900 & -0.243 & 182 \\
  26-12-2018 & 58478.6664 & 2.4 m/YFOSC  & 4970-9830 &  0.061 & 267 \\
  03-11-2019 & 58790.8476 & 2.16 m/OMR   & 5500-6900 &  0.372 & 191 \\
  26-12-2019 & 58843.7108 & 2.4 m/YFOSC  & 4970-9830 &  0.113 & 239 \\
  20-02-2020 & 58900.0960 & 0.7 m/AZT-8  & 6100-6800 &  0.176 & 183 \\
  09-10-2020 & 59131.8682 & 2.16 m/BFOSC & 5950-8200 &  0.303 & 100 \\
  10-10-2020 & 59132.6998 & 2.16 m/BFOSC & 5950-8200 &  0.226 & 139 \\
  11-10-2020 & 59133.8590 & 2.16 m/BFOSC & 5950-8200 &  0.028 & 176 \\
  12-10-2020 & 59134.7982 & 2.16 m/BFOSC & 5950-8200 &  0.097 & 145 \\
  24-11-2020 & 59177.6292 & 2.16 m/BFOSC & 5950-8200 &  0.411 & 176 \\
  24-11-2020 & 59177.6442 & 2.16 m/BFOSC & 5950-8200 &  0.174 & 158 \\
  24-11-2020 & 59177.6591 & 2.16 m/BFOSC & 5950-8200 &  0.291 & 163 \\
  25-11-2020 & 59178.7197 & 2.16 m/BFOSC & 5950-8200 &  0.145 & 165 \\
  25-11-2020 & 59178.7408 & 2.16 m/BFOSC & 5950-8200 &  0.120 & 173 \\
  26-11-2020 & 59179.8182 & 2.16 m/BFOSC & 5950-8200 &  0.074 & 154 \\
  27-11-2020 & 59180.7689 & 2.16 m/BFOSC & 5950-8200 &  0.312 & 139 \\
  27-11-2020 & 59180.7798 & 2.16 m/BFOSC & 5950-8200 &  0.131 & 180 \\
  27-11-2020 & 59180.7868 & 2.16 m/BFOSC & 5950-8200 &  0.133 & 180 \\
  27-11-2020 & 59180.7938 & 2.16 m/BFOSC & 5950-8200 &  0.026 & 181 \\
  27-11-2020 & 59180.8008 & 2.16 m/BFOSC & 5950-8200 &  0.207 & 170 \\
  28-11-2020 & 59181.8820 & 2.16 m/BFOSC & 5950-8200 &  0.303 & 145 \\
  28-11-2020 & 59181.8893 & 2.16 m/BFOSC & 5950-8200 & -0.045 & 180 \\
  28-11-2020 & 59181.8928 & 2.16 m/BFOSC & 5950-8200 &  0.362 & 155 \\
  28-11-2020 & 59181.8963 & 2.16 m/BFOSC & 5950-8200 &  0.384 & 136 \\
  28-11-2020 & 59181.8999 & 2.16 m/BFOSC & 5950-8200 &  0.284 & 175 \\
  28-11-2020 & 59181.9034 & 2.16 m/BFOSC & 5950-8200 &  0.077 & 183 \\
  28-11-2020 & 59181.9070 & 2.16 m/BFOSC & 5950-8200 &  0.145 & 172 \\
  28-11-2020 & 59181.9105 & 2.16 m/BFOSC & 5950-8200 &  0.042 & 126 \\
  28-11-2020 & 59181.9140 & 2.16 m/BFOSC & 5950-8200 &  0.208 & 165 \\
  29-11-2020 & 59182.8790 & 2.16 m/BFOSC & 5950-8200 & -0.064 & 204 \\
  29-11-2020 & 59182.8825 & 2.16 m/BFOSC & 5950-8200 & -0.058 & 201 \\
  29-11-2020 & 59182.8860 & 2.16 m/BFOSC & 5950-8200 &  0.043 & 193 \\
  29-11-2020 & 59182.8896 & 2.16 m/BFOSC & 5950-8200 & -0.041 & 197 \\
  29-11-2020 & 59182.8931 & 2.16 m/BFOSC & 5950-8200 &  0.073 & 196 \\
  29-11-2020 & 59182.8967 & 2.16 m/BFOSC & 5950-8200 &  0.036 & 195 \\
  29-11-2020 & 59182.9002 & 2.16 m/BFOSC & 5950-8200 &  0.050 & 196 \\
  29-11-2020 & 59182.9037 & 2.16 m/BFOSC & 5950-8200 & -0.008 & 200 \\
  29-11-2020 & 59182.9073 & 2.16 m/BFOSC & 5950-8200 & -0.055 & 194 \\
  29-11-2020 & 59182.9108 & 2.16 m/BFOSC & 5950-8200 &  0.036 & 187 \\
  10-10-2021 & 59497.7769 & 2.16 m/OMR   & 6000-6850 &  0.264 & 224 \\
\end{longtable}

\section{Table of photometric observations}

\begin{longtable}{cccccc}
\caption{Summary of photometric observations of 1A 0535+26.\label{sec:table2}}\\
\hline 
\hline 
 MJD & Telescope & B(mag) & V(mag) & R(mag) & I(mag) \\
\hline
\endfirsthead
\caption{Continued.} \\
\hline 
\hline 
 MJD & Telescope & B(mag) & V(mag) & R(mag) & I(mag) \\
\hline
\endhead
\hline
\endfoot
\hline
\endlastfoot
55481.7577 & 60cm & 9.791 $\pm$ 0.014 & 9.353 $\pm$ 0.011 & 9.000 $\pm$ 0.010 & 8.558 $\pm$ 0.018 \\
55482.8444 & 80cm & 9.764 $\pm$ 0.014 & 9.328 $\pm$ 0.011 & 8.972 $\pm$ 0.010 & 8.537 $\pm$ 0.018 \\
55483.7726 & 80cm & 9.767 $\pm$ 0.014 & 9.339 $\pm$ 0.010 & 8.982 $\pm$ 0.011 & 8.530 $\pm$ 0.018 \\
55484.7981 & 80cm & 9.784 $\pm$ 0.014 & 9.363 $\pm$ 0.011 & 8.993 $\pm$ 0.010 & 8.535 $\pm$ 0.018 \\
55822.8024 & 80cm & 9.876 $\pm$ 0.014 & 9.437 $\pm$ 0.010 & 9.091 $\pm$ 0.010 & 8.702 $\pm$ 0.018 \\
56001.6156 & 2.4m &       ...         & 9.177 $\pm$ 0.010 &       ...         &       ...         \\
56002.5662 & 2.4m & 9.538 $\pm$ 0.014 & 9.170 $\pm$ 0.010 &       ...         &       ...         \\
56223.7232 & 60cm & 9.591 $\pm$ 0.014 & 9.027 $\pm$ 0.010 & 8.614 $\pm$ 0.010 & 8.096 $\pm$ 0.018 \\
56556.7710 & 80cm & 9.687 $\pm$ 0.014 &       ...         & 8.780 $\pm$ 0.010 & 8.287 $\pm$ 0.018 \\
56558.8179 & 80cm & 9.668 $\pm$ 0.014 & 9.153 $\pm$ 0.010 & 8.766 $\pm$ 0.010 & 8.243 $\pm$ 0.018 \\
56590.7604 & 80cm & 9.669 $\pm$ 0.014 & 9.176 $\pm$ 0.010 & 8.799 $\pm$ 0.010 &       ...         \\
56591.7473 & 80cm & 9.676 $\pm$ 0.014 & 9.158 $\pm$ 0.010 & 8.786 $\pm$ 0.010 & 8.268 $\pm$ 0.018 \\
56592.6428 & 80cm & 9.667 $\pm$ 0.014 & 9.146 $\pm$ 0.010 & 8.787 $\pm$ 0.010 & 8.295 $\pm$ 0.018 \\
56594.8470 & 80cm & 9.681 $\pm$ 0.014 & 9.175 $\pm$ 0.010 & 8.801 $\pm$ 0.010 & 8.295 $\pm$ 0.018 \\
56595.6518 & 80cm & 9.689 $\pm$ 0.014 & 9.187 $\pm$ 0.010 & 8.813 $\pm$ 0.010 & 8.299 $\pm$ 0.018 \\
56596.6759 & 80cm & 9.698 $\pm$ 0.014 & 9.185 $\pm$ 0.010 & 8.805 $\pm$ 0.010 & 8.313 $\pm$ 0.018 \\
56664.6378 & 60cm & 9.700 $\pm$ 0.014 & 9.196 $\pm$ 0.010 & 8.793 $\pm$ 0.010 & 8.302 $\pm$ 0.018 \\
56665.6980 & 60cm & 9.650 $\pm$ 0.014 & 9.194 $\pm$ 0.011 & 8.793 $\pm$ 0.011 & 8.293 $\pm$ 0.018 \\
56666.7638 & 60cm & 9.662 $\pm$ 0.014 & 9.187 $\pm$ 0.011 & 8.761 $\pm$ 0.011 & 8.304 $\pm$ 0.018 \\
56667.7075 & 60cm & 9.639 $\pm$ 0.014 & 9.161 $\pm$ 0.012 & 8.745 $\pm$ 0.011 & 8.262 $\pm$ 0.018 \\
56685.5394 & 60cm & 9.682 $\pm$ 0.014 & 9.152 $\pm$ 0.010 & 8.808 $\pm$ 0.010 & 8.336 $\pm$ 0.018 \\
56720.4742 & 60cm & 9.699 $\pm$ 0.014 & 9.178 $\pm$ 0.010 & 8.859 $\pm$ 0.010 & 8.376 $\pm$ 0.018 \\
56722.4760 & 60cm & 9.695 $\pm$ 0.014 & 9.190 $\pm$ 0.010 & 8.839 $\pm$ 0.010 & 8.361 $\pm$ 0.018 \\
56726.4769 & 60cm &       ...         & 9.213 $\pm$ 0.010 &       ...         & 8.398 $\pm$ 0.018 \\
56917.8152 & 60cm &       ...         & 9.151 $\pm$ 0.011 &       ...         &       ...         \\
56926.8108 & 60cm & 9.683 $\pm$ 0.014 & 9.168 $\pm$ 0.010 & 8.815 $\pm$ 0.010 & 8.338 $\pm$ 0.018 \\
56928.8289 & 60cm & 9.713 $\pm$ 0.014 & 9.132 $\pm$ 0.010 & 8.889 $\pm$ 0.018 & 8.296 $\pm$ 0.018 \\
56929.8003 & 80cm & 9.694 $\pm$ 0.014 & 9.161 $\pm$ 0.010 & 8.818 $\pm$ 0.010 & 8.327 $\pm$ 0.018 \\
57045.5445 & 80cm & 9.741 $\pm$ 0.014 & 9.299 $\pm$ 0.010 & 8.839 $\pm$ 0.010 & 8.437 $\pm$ 0.018 \\
57048.5984 & 80cm & 9.571 $\pm$ 0.014 & 9.076 $\pm$ 0.010 & 8.750 $\pm$ 0.010 & 8.288 $\pm$ 0.018 \\
57296.8892 & 80cm & 9.740 $\pm$ 0.014 & 9.396 $\pm$ 0.010 & 8.840 $\pm$ 0.010 & 8.368 $\pm$ 0.018 \\
57297.8045 & 80cm & 9.564 $\pm$ 0.014 & 9.067 $\pm$ 0.010 & 8.684 $\pm$ 0.010 & 8.194 $\pm$ 0.018 \\
57299.7901 & 80cm & 9.549 $\pm$ 0.014 & 9.168 $\pm$ 0.010 & 8.658 $\pm$ 0.010 & 8.189 $\pm$ 0.018 \\
57300.7396 & 80cm & 9.542 $\pm$ 0.014 & 9.107 $\pm$ 0.010 & 8.668 $\pm$ 0.010 & 8.175 $\pm$ 0.018 \\
57303.7336 & 80cm & 9.602 $\pm$ 0.014 & 9.056 $\pm$ 0.010 & 8.709 $\pm$ 0.010 & 8.222 $\pm$ 0.018 \\
57355.7953 & 80cm & 9.510 $\pm$ 0.014 & 8.976 $\pm$ 0.010 & 8.626 $\pm$ 0.010 & 8.085 $\pm$ 0.018 \\
57356.5800 & 80cm & 9.513 $\pm$ 0.014 & 8.991 $\pm$ 0.010 & 8.620 $\pm$ 0.010 & 8.104 $\pm$ 0.018 \\
57650.7584 & 60cm & 9.470 $\pm$ 0.014 & 8.937 $\pm$ 0.011 & 8.469 $\pm$ 0.011 & 7.936 $\pm$ 0.018 \\
57651.7032 & 60cm & 9.463 $\pm$ 0.015 & 8.968 $\pm$ 0.012 & 8.504 $\pm$ 0.011 & 7.941 $\pm$ 0.018 \\
57658.8160 & 60cm & 9.485 $\pm$ 0.014 & 8.937 $\pm$ 0.010 & 8.495 $\pm$ 0.010 & 7.984 $\pm$ 0.018 \\
57659.8562 & 60cm & 9.471 $\pm$ 0.014 & 8.927 $\pm$ 0.011 & 8.485 $\pm$ 0.011 & 7.981 $\pm$ 0.018 \\
57684.8813 & 60cm & 9.453 $\pm$ 0.014 & 8.913 $\pm$ 0.011 & 8.477 $\pm$ 0.011 & 7.926 $\pm$ 0.020 \\
57685.8649 & 60cm & 9.468 $\pm$ 0.014 & 8.927 $\pm$ 0.011 & 8.466 $\pm$ 0.012 & 7.910 $\pm$ 0.020 \\
57686.8695 & 60cm & 9.488 $\pm$ 0.014 & 8.917 $\pm$ 0.012 & 8.473 $\pm$ 0.013 & 7.935 $\pm$ 0.019 \\
57689.9000 & 60cm & 9.481 $\pm$ 0.014 & 8.934 $\pm$ 0.011 & 8.462 $\pm$ 0.013 & 7.920 $\pm$ 0.021 \\
57690.7980 & 60cm & 9.472 $\pm$ 0.014 & 8.927 $\pm$ 0.012 & 8.475 $\pm$ 0.012 & 7.923 $\pm$ 0.021 \\
57691.8255 & 60cm & 9.476 $\pm$ 0.014 & 8.940 $\pm$ 0.012 & 8.483 $\pm$ 0.011 & 7.952 $\pm$ 0.018 \\
57692.7812 & 60cm & 9.471 $\pm$ 0.015 & 8.933 $\pm$ 0.011 & 8.477 $\pm$ 0.011 & 7.922 $\pm$ 0.020 \\
57714.7961 & 60cm & 9.446 $\pm$ 0.014 & 8.937 $\pm$ 0.010 & 8.414 $\pm$ 0.011 & 7.930 $\pm$ 0.018 \\
57715.6319 & 60cm & 9.477 $\pm$ 0.014 & 8.940 $\pm$ 0.010 & 8.478 $\pm$ 0.011 & 7.936 $\pm$ 0.018 \\
57716.8733 & 60cm & 9.469 $\pm$ 0.014 & 8.936 $\pm$ 0.011 & 8.466 $\pm$ 0.012 & 7.921 $\pm$ 0.020 \\
57718.7832 & 60cm & 9.488 $\pm$ 0.014 & 8.938 $\pm$ 0.011 & 8.463 $\pm$ 0.013 & 7.929 $\pm$ 0.024 \\
57719.8161 & 60cm & 9.483 $\pm$ 0.014 & 8.944 $\pm$ 0.011 & 8.479 $\pm$ 0.010 & 7.954 $\pm$ 0.018 \\
57720.6421 & 60cm & 9.462 $\pm$ 0.014 & 8.934 $\pm$ 0.011 & 8.463 $\pm$ 0.011 & 7.933 $\pm$ 0.019 \\
57722.6457 & 60cm & 9.463 $\pm$ 0.014 & 8.932 $\pm$ 0.010 & 8.479 $\pm$ 0.010 & 7.939 $\pm$ 0.018 \\
57723.6104 & 60cm & 9.474 $\pm$ 0.014 & 8.947 $\pm$ 0.010 & 8.495 $\pm$ 0.011 & 7.938 $\pm$ 0.018 \\
57744.6336 & 60cm & 9.491 $\pm$ 0.014 & 8.946 $\pm$ 0.010 & 8.498 $\pm$ 0.010 & 7.959 $\pm$ 0.018 \\
57745.6188 & 60cm & 9.471 $\pm$ 0.014 & 8.938 $\pm$ 0.010 & 8.483 $\pm$ 0.011 & 7.946 $\pm$ 0.019 \\
57746.7854 & 60cm & 9.459 $\pm$ 0.014 & 8.922 $\pm$ 0.010 & 8.453 $\pm$ 0.011 & 7.922 $\pm$ 0.019 \\
57748.7089 & 60cm & 9.507 $\pm$ 0.014 & 8.931 $\pm$ 0.012 & 8.477 $\pm$ 0.014 & 7.930 $\pm$ 0.022 \\
57749.7842 & 60cm & 9.482 $\pm$ 0.014 & 8.930 $\pm$ 0.011 & 8.463 $\pm$ 0.013 & 7.927 $\pm$ 0.019 \\
57750.6546 & 60cm & 9.488 $\pm$ 0.014 & 8.927 $\pm$ 0.010 & 8.463 $\pm$ 0.011 & 7.925 $\pm$ 0.018 \\
57751.5938 & 60cm & 9.483 $\pm$ 0.014 & 8.932 $\pm$ 0.010 & 8.456 $\pm$ 0.011 & 7.904 $\pm$ 0.019 \\
57752.7125 & 60cm & 9.484 $\pm$ 0.014 & 8.934 $\pm$ 0.010 & 8.461 $\pm$ 0.011 & 7.901 $\pm$ 0.018 \\
57753.6965 & 60cm & 9.472 $\pm$ 0.014 & 8.930 $\pm$ 0.010 & 8.466 $\pm$ 0.011 & 7.942 $\pm$ 0.019 \\
57754.7377 & 60cm & 9.479 $\pm$ 0.014 & 8.931 $\pm$ 0.010 & 8.463 $\pm$ 0.011 & 7.926 $\pm$ 0.018 \\
57755.7211 & 60cm & 9.475 $\pm$ 0.014 & 8.938 $\pm$ 0.010 & 8.464 $\pm$ 0.011 & 7.935 $\pm$ 0.019 \\
57756.6092 & 60cm & 9.499 $\pm$ 0.014 & 8.937 $\pm$ 0.010 & 8.469 $\pm$ 0.011 & 7.921 $\pm$ 0.018 \\
57757.6852 & 60cm & 9.459 $\pm$ 0.017 & 8.931 $\pm$ 0.011 & 8.460 $\pm$ 0.011 & 7.920 $\pm$ 0.019 \\
57758.7361 & 60cm & 9.447 $\pm$ 0.015 & 8.927 $\pm$ 0.011 & 8.465 $\pm$ 0.011 & 7.919 $\pm$ 0.019 \\
57759.6367 & 60cm & 9.477 $\pm$ 0.015 & 8.918 $\pm$ 0.011 & 8.456 $\pm$ 0.011 & 7.917 $\pm$ 0.019 \\
57761.6171 & 60cm & 9.500 $\pm$ 0.014 & 8.943 $\pm$ 0.010 & 8.466 $\pm$ 0.011 & 7.935 $\pm$ 0.018 \\
57762.6608 & 60cm & 9.492 $\pm$ 0.015 & 8.938 $\pm$ 0.011 & 8.466 $\pm$ 0.011 & 7.923 $\pm$ 0.019 \\
57802.5468 & 60cm & 9.481 $\pm$ 0.015 & 8.911 $\pm$ 0.012 & 8.491 $\pm$ 0.014 & 7.939 $\pm$ 0.024 \\
57806.6700 & 60cm &       ...         & 8.946 $\pm$ 0.010 & 8.473 $\pm$ 0.011 & 7.945 $\pm$ 0.018 \\
57807.6713 & 60cm &       ...         & 8.918 $\pm$ 0.010 & 8.472 $\pm$ 0.011 & 7.923 $\pm$ 0.018 \\
57808.5487 & 60cm &       ...         & 8.931 $\pm$ 0.010 & 8.456 $\pm$ 0.011 & 7.965 $\pm$ 0.018 \\
57809.5464 & 60cm &       ...         & 8.945 $\pm$ 0.010 & 8.465 $\pm$ 0.011 & 7.940 $\pm$ 0.018 \\
57810.5369 & 60cm &       ...         & 8.950 $\pm$ 0.010 & 8.456 $\pm$ 0.011 &       ...         \\
57811.5758 & 60cm &       ...         & 8.926 $\pm$ 0.011 & 8.459 $\pm$ 0.012 & 7.923 $\pm$ 0.018 \\
58012.7512 & 60cm & 9.506 $\pm$ 0.014 & 8.961 $\pm$ 0.010 & 8.464 $\pm$ 0.011 & 7.908 $\pm$ 0.019 \\
58013.8090 & 60cm & 9.485 $\pm$ 0.014 & 8.948 $\pm$ 0.010 & 8.427 $\pm$ 0.011 & 7.902 $\pm$ 0.019 \\
58014.7710 & 60cm & 9.495 $\pm$ 0.014 & 8.916 $\pm$ 0.010 & 8.436 $\pm$ 0.011 & 7.921 $\pm$ 0.018 \\
58015.8748 & 60cm & 9.508 $\pm$ 0.014 & 8.954 $\pm$ 0.010 & 8.455 $\pm$ 0.011 & 7.907 $\pm$ 0.018 \\
58053.8587 & 60cm & 9.537 $\pm$ 0.014 & 8.970 $\pm$ 0.011 & 8.481 $\pm$ 0.011 & 7.920 $\pm$ 0.019 \\
58054.8382 & 60cm & 9.526 $\pm$ 0.014 & 8.970 $\pm$ 0.010 & 8.482 $\pm$ 0.011 & 7.921 $\pm$ 0.018 \\
58055.6948 & 60cm & 9.524 $\pm$ 0.014 & 8.978 $\pm$ 0.010 & 8.485 $\pm$ 0.011 & 7.913 $\pm$ 0.018 \\
58056.7002 & 60cm & 9.516 $\pm$ 0.014 &       ...         & 8.477 $\pm$ 0.011 & 7.916 $\pm$ 0.019 \\
58061.7369 & 60cm & 9.503 $\pm$ 0.014 & 8.972 $\pm$ 0.011 & 8.480 $\pm$ 0.011 & 7.907 $\pm$ 0.019 \\
58062.6791 & 60cm & 9.509 $\pm$ 0.015 & 8.950 $\pm$ 0.012 & 8.492 $\pm$ 0.012 & 7.912 $\pm$ 0.020 \\
58064.6826 & 60cm & 9.496 $\pm$ 0.014 & 8.959 $\pm$ 0.010 & 8.494 $\pm$ 0.011 & 7.923 $\pm$ 0.019 \\
58065.7206 & 60cm & 9.506 $\pm$ 0.015 & 8.951 $\pm$ 0.012 & 8.469 $\pm$ 0.012 & 7.920 $\pm$ 0.020 \\
58075.8819 & 60cm & 9.482 $\pm$ 0.014 & 8.926 $\pm$ 0.010 & 8.463 $\pm$ 0.011 & 7.929 $\pm$ 0.018 \\
58076.8941 & 60cm & 9.485 $\pm$ 0.014 & 8.944 $\pm$ 0.010 & 8.476 $\pm$ 0.011 & 7.936 $\pm$ 0.019 \\
58077.8788 & 60cm & 9.482 $\pm$ 0.015 & 8.950 $\pm$ 0.011 & 8.469 $\pm$ 0.011 & 7.936 $\pm$ 0.019 \\
58078.8631 & 60cm & 9.490 $\pm$ 0.015 & 8.943 $\pm$ 0.011 & 8.475 $\pm$ 0.011 & 7.916 $\pm$ 0.019 \\
58079.8484 & 60cm & 9.506 $\pm$ 0.014 & 8.959 $\pm$ 0.011 & 8.475 $\pm$ 0.011 & 7.913 $\pm$ 0.018 \\
58080.8131 & 60cm & 9.499 $\pm$ 0.014 & 8.955 $\pm$ 0.011 & 8.464 $\pm$ 0.011 & 7.915 $\pm$ 0.018 \\
58083.5024 & 60cm & 9.498 $\pm$ 0.014 & 8.959 $\pm$ 0.011 & 8.464 $\pm$ 0.011 & 7.919 $\pm$ 0.019 \\
58085.5187 & 60cm & 9.517 $\pm$ 0.014 & 8.969 $\pm$ 0.010 & 8.495 $\pm$ 0.011 & 7.932 $\pm$ 0.019 \\
58086.6695 & 60cm & 9.508 $\pm$ 0.014 & 8.972 $\pm$ 0.011 & 8.479 $\pm$ 0.012 & 7.928 $\pm$ 0.020 \\
58087.6657 & 60cm & 9.510 $\pm$ 0.014 & 8.966 $\pm$ 0.010 & 8.476 $\pm$ 0.011 & 7.937 $\pm$ 0.018 \\
58088.7191 & 60cm & 9.519 $\pm$ 0.014 & 8.975 $\pm$ 0.011 & 8.488 $\pm$ 0.011 & 7.919 $\pm$ 0.019 \\
58089.6599 & 60cm & 9.524 $\pm$ 0.015 & 8.972 $\pm$ 0.011 & 8.475 $\pm$ 0.012 & 7.937 $\pm$ 0.019 \\
58090.7504 & 60cm & 9.517 $\pm$ 0.015 & 8.972 $\pm$ 0.011 & 8.495 $\pm$ 0.012 & 7.945 $\pm$ 0.020 \\
58091.7591 & 60cm & 9.522 $\pm$ 0.014 & 8.972 $\pm$ 0.011 & 8.484 $\pm$ 0.011 & 7.923 $\pm$ 0.019 \\
58092.5340 & 60cm & 9.521 $\pm$ 0.015 & 8.974 $\pm$ 0.012 & 8.498 $\pm$ 0.012 & 7.948 $\pm$ 0.019 \\
58093.6669 & 60cm & 9.534 $\pm$ 0.017 & 8.959 $\pm$ 0.011 & 8.483 $\pm$ 0.011 & 7.930 $\pm$ 0.019 \\
58094.7367 & 60cm & 9.514 $\pm$ 0.014 & 8.967 $\pm$ 0.011 & 8.485 $\pm$ 0.011 & 7.923 $\pm$ 0.019 \\
58117.5499 & 60cm & 9.518 $\pm$ 0.014 & 8.972 $\pm$ 0.010 & 8.482 $\pm$ 0.011 & 7.913 $\pm$ 0.019 \\
58118.5161 & 60cm & 9.520 $\pm$ 0.014 & 8.969 $\pm$ 0.010 & 8.475 $\pm$ 0.011 & 7.917 $\pm$ 0.019 \\
58134.6746 & 60cm & 9.530 $\pm$ 0.014 & 8.969 $\pm$ 0.011 & 8.488 $\pm$ 0.011 & 7.914 $\pm$ 0.019 \\
58135.6052 & 60cm & 9.510 $\pm$ 0.014 & 8.969 $\pm$ 0.010 & 8.485 $\pm$ 0.011 & 7.907 $\pm$ 0.019 \\
58136.6026 & 60cm & 9.514 $\pm$ 0.016 & 8.981 $\pm$ 0.015 & 8.505 $\pm$ 0.016 & 7.859 $\pm$ 0.025 \\
58149.5716 & 60cm & 9.500 $\pm$ 0.014 & 8.956 $\pm$ 0.010 & 8.458 $\pm$ 0.011 & 7.888 $\pm$ 0.019 \\
58151.6266 & 60cm & 9.527 $\pm$ 0.014 & 8.969 $\pm$ 0.010 & 8.489 $\pm$ 0.011 & 7.922 $\pm$ 0.019 \\
58153.6702 & 60cm & 9.539 $\pm$ 0.015 & 8.994 $\pm$ 0.011 & 8.496 $\pm$ 0.011 & 7.895 $\pm$ 0.019 \\
58370.8243 & 80cm & 9.509 $\pm$ 0.015 & 8.985 $\pm$ 0.011 & 8.507 $\pm$ 0.012 & 7.975 $\pm$ 0.020 \\
58377.8316 & 80cm & 9.513 $\pm$ 0.014 & 8.986 $\pm$ 0.010 & 8.511 $\pm$ 0.011 & 7.966 $\pm$ 0.019 \\
58378.8528 & 80cm & 9.545 $\pm$ 0.014 & 8.988 $\pm$ 0.010 & 8.514 $\pm$ 0.012 & 7.994 $\pm$ 0.019 \\
58428.8303 & 80cm & 9.507 $\pm$ 0.014 & 8.992 $\pm$ 0.010 & 8.536 $\pm$ 0.011 & 7.997 $\pm$ 0.019 \\
58430.9038 & 80cm & 9.524 $\pm$ 0.014 & 9.008 $\pm$ 0.011 & 8.560 $\pm$ 0.011 & 8.009 $\pm$ 0.019 \\
58456.7084 & 80cm & 9.471 $\pm$ 0.014 & 8.961 $\pm$ 0.011 & 8.524 $\pm$ 0.011 & 7.984 $\pm$ 0.019 \\
58750.8628 & 80cm & 9.555 $\pm$ 0.014 & 9.060 $\pm$ 0.011 & 8.618 $\pm$ 0.012 & 8.093 $\pm$ 0.021 \\
58751.8327 & 80cm & 9.588 $\pm$ 0.014 & 9.084 $\pm$ 0.012 & 8.618 $\pm$ 0.012 & 8.134 $\pm$ 0.019 \\
58760.8349 & 80cm & 9.586 $\pm$ 0.014 & 9.097 $\pm$ 0.011 & 8.661 $\pm$ 0.011 & 8.146 $\pm$ 0.019 \\
58790.8072 & 80cm & 9.564 $\pm$ 0.014 & 9.078 $\pm$ 0.012 & 8.672 $\pm$ 0.011 & 8.180 $\pm$ 0.019 \\
58791.8955 & 80cm & 9.575 $\pm$ 0.014 & 9.108 $\pm$ 0.011 & 8.677 $\pm$ 0.011 & 8.158 $\pm$ 0.019 \\
58793.8840 & 80cm & 9.565 $\pm$ 0.014 & 9.088 $\pm$ 0.011 & 8.672 $\pm$ 0.011 & 8.200 $\pm$ 0.019 \\
59107.7856 & 80cm & 9.713 $\pm$ 0.015 & 9.324 $\pm$ 0.012 & 8.945 $\pm$ 0.012 & 8.519 $\pm$ 0.020 \\
59108.8351 & 80cm & 9.717 $\pm$ 0.015 & 9.314 $\pm$ 0.011 & 8.921 $\pm$ 0.012 & 8.520 $\pm$ 0.020 \\
59131.8483 & 80cm & 9.739 $\pm$ 0.017 & 9.321 $\pm$ 0.014 & 8.978 $\pm$ 0.014 & 8.540 $\pm$ 0.022 \\
59133.8667 & 80cm & 9.775 $\pm$ 0.015 & 9.375 $\pm$ 0.011 & 9.023 $\pm$ 0.012 & 8.592 $\pm$ 0.020 \\
59173.6600 & 80cm & 9.765 $\pm$ 0.014 & 9.360 $\pm$ 0.010 & 9.002 $\pm$ 0.010 & 8.592 $\pm$ 0.018 \\
59175.7750 & 80cm & 9.753 $\pm$ 0.014 & 9.352 $\pm$ 0.010 & 9.006 $\pm$ 0.010 & 8.585 $\pm$ 0.018 \\
59177.7300 & 80cm & 9.747 $\pm$ 0.014 & 9.346 $\pm$ 0.010 & 9.008 $\pm$ 0.010 & 8.605 $\pm$ 0.018 \\
59201.7036 & 80cm & 9.797 $\pm$ 0.015 & 9.399 $\pm$ 0.012 & 9.059 $\pm$ 0.012 & 8.659 $\pm$ 0.020 \\
59202.5830 & 80cm &       ...         & 9.400 $\pm$ 0.012 & 9.031 $\pm$ 0.012 & 8.633 $\pm$ 0.020 \\
59203.7075 & 80cm & 9.801 $\pm$ 0.015 & 9.395 $\pm$ 0.012 & 9.056 $\pm$ 0.012 & 8.628 $\pm$ 0.020 \\
59204.7810 & 80cm & 9.798 $\pm$ 0.015 & 9.398 $\pm$ 0.012 & 9.049 $\pm$ 0.012 & 8.633 $\pm$ 0.021 \\
59497.8512 & 80cm & 9.799 $\pm$ 0.015 & 9.427 $\pm$ 0.012 & 9.050 $\pm$ 0.012 & 8.649 $\pm$ 0.020 \\
\end{longtable}

\end{appendix}

\end{document}